\documentstyle[12pt]{article}

\newcommand{\beq}[1]{\begin{equation}\label{#1}}
\newcommand{\eeq}{\end{equation}}
\newcommand{\bear}[1]{\begin{eqnarray}\label{#1}}
\newcommand{\ear}{\end{eqnarray}}
\newcommand{\nn}{\nonumber}
\newcommand{\rf}[1]{(\ref{#1})}

\catcode`\@=11 \@addtoreset{equation}{section}\catcode`\@=12
\newcommand{\be}{\begin{equation}}
\newcommand{\ee}{\end{equation}}
\newcommand{\ba}{\begin{eqnarray}}
\newcommand{\ea}{\end{eqnarray}}

\newcommand{\np}{ {\newpage } }

\newcommand{\Iff}{ {\Leftrightarrow } }

\newcommand{\N}{ \mbox{\rm I$\!$N} }
\newcommand{\R}{ \mbox{\rm I$\!$R} }
\def\C{\mbox{\rm {I\kern-.520em C}}}

\newcommand{\sign}{ \mbox{\rm sign} }
\newcommand{\e}{ \mbox{\rm e} }
\newcommand{\eps}{ \varepsilon }

\newcommand{\rank}{ \mbox{\rm rank} }

\newcommand{\p}{\partial}
\newcommand{\btd}{\bigtriangledown}
\newcommand{\btu}{\bigtriangleup}
\newcommand{\tri}{\Delta}
\newcommand{\sq}[1]{\sqrt{|#1|}}
\newcommand{\avers}[1]{{<{#1}>}_*}
\newcommand{\til}[1]{\tilde{#1}}

\begin{document}

\centerline{\large \bf
Sigma-model for Generalized  Composite p-branes}

\vspace{1.03truecm}

\bigskip

\centerline{\bf \large
V. D. Ivashchuk and V. N. Melnikov}

\vspace{0.96truecm}

\centerline{Center for Gravitation and Fundamental Metrology}
\centerline{VNIIMS, 3-1 M. Ulyanovoy Str.}
\centerline{Moscow, 117313, Russia}
\centerline{e-mail: melnikov@fund.phys.msu.su}

\begin{abstract}

A multidimensional gravitational model
containing  several dilatonic scalar fields
and antisymmetric forms is considered. The manifold
is chosen in the form $M = M_0 \times M_1 \times \ldots
\times M_n$,
where  $M_i$ are Einstein spaces ($i \geq 1$).
The block-diagonal metric is chosen and all fields and scale
factors of the metric are functions on $M_0$.
For the forms composite (electro-magnetic) p-brane ansatz is adopted.
The model is reduced to gravitating self-interacting sigma-model
with certain constraints.
In pure electric and magnetic
cases the number of these constraints is $n_1(n_1 - 1)/2$
where $n_1$ is number of 1-dimensional manifolds among $M_i$.
In the "electro-magnetic" case for ${\rm dim} M_0 = 1, 3$ additional
$n_1$ constraints appear.
A family of "Majumdar-Papapetrou type"  solutions
governed by a set of harmonic functions
is obtained, when all factor-spaces $M_{\nu}$ are Ricci-flat.
These solutions are generalized to the case of
non-Ricci-flat $M_0$ when also some additional
"internal" Einstein spaces of non-zero curvature are
added to $M$.  As an example
exact solutions for  $D = 11$ supergravity and
related 12-dimensional theory are presented.

\end{abstract}
\hspace*{0.950cm} PACS number(s):\ 04.50.+h,\ 98.80.Hw,\ 04.60.Kz
\np

\section{\bf Introduction}
\setcounter{equation}{0}

At present there exists an interest in  studying of  so-called
$p$-brane solutions in multidimensional models \cite{Dab}-\cite{IMR}
(for a review see \cite{St}). These solutions generalize well-known
Majumdar-Papapetrou solutions \cite{MP,Br} to the case when several
antisymmetric forms and dilatonic scalar fields are considered.

In this paper we continue our investigation
of p-brane solutions \cite{IM4,IMR} based on the
$\sigma$-model approach to the composite electro-magnetic case, i.e. when
antisymmetric forms are sums of elementary solutions. Solutions of such
type are objects of intensive investigations in $D=10,11$
supergravities \cite{CJS,SS} (and in theory of superstrings and $M$-theory
\cite{GSW,HTW,S,D}).

Here we obtain the $\sigma$-model representation for the
composite electro-magnetic p-brane ansatz in the
multidimensional gravitational model with several scalar
fields and antisymmetric fields (forms). The manifold is chosen
in the form $M = M_0 \times M_1 \times \ldots \times M_n$,
where  $M_i$ are Einstein spaces ($i \geq 1$).  In opposite to
non-composite case  \cite{IM4} here a set of constraints on
$\sigma$-model fields appears. These constraints occur due to
non-diagonality of the stress-energy tensor $T^M_N$.

For Ricci-flat $M_i$, $i \geq 1$, we obtain
>from the $\sigma$-model representation  a set of
Majumdar-Papapetrou-type solutions,
using the relations for scalar products
of some vectors in "midisuperspace" metric. Thus, here like
in \cite{IM4} we extend our
approach used in multidimensional cosmology (based on reduction to
Toda-like systems) to $\sigma$-models of special type
generated by interacting scalar, gravitational fields
and generalized electro-magnetic fields (forms).
For flat $M_i$, $i \geq 1$, in certain special
cases our solutions agree with  those  considered
earlier (see, for example \cite{AR,AEH,AIR} and
references  therein). The solutions are presented in the form
generalizing harmonic function rule of     \cite{Ts1}
(see also \cite{PT,GKT}).

Here we generalize the obtained solutions to the case of
non-Ricci-flat $M_0$ when also some
"internal" Einstein spaces of non-zero curvature are also
added. In this case  we "split"  finding  exact solutions
in two parts: (i) first we  solve the "background"
field equations for the metric and scalar fields obeying
certain "p-brane restrictions" and then (ii) we construct
the generalized intersecting p-brane solutions governed
by a set of harmonic functions on $M_0$ with the background metric.

We also note that here we start with a metric of arbitrary
signature. This may have applications for supergravitational models
with several times \cite{HP,N}. Also,
we consider a scalar field  kinetic term of arbitrary signature
(such situation takes place, for example, in 12-dimensional
model from     \cite{KKP} that may
correspond to F-theory \cite{Vafa}).

\section{\bf The model}
\setcounter{equation}{0}

We consider the model governed by an action
\bear{2.1}
S =&& \frac{1}{2\kappa^{2}}
\int_{M} d^{D}z \sqrt{|g|} \{ {R}[g] - 2 \Lambda - C_{\alpha\beta}
g^{MN} \partial_{M} \varphi^\alpha \partial_{N} \varphi^\beta
\\ \nn
&& - \sum_{a \in \Delta}
\frac{\theta_a}{n_a!} \exp[ 2 \lambda_{a} (\varphi) ] (F^a)^2_g \}
+ S_{GH},
\ear
where $g = g_{MN} dz^{M} \otimes dz^{N}$ is the metric
($M,N =1, \ldots, D$),
$\varphi=(\varphi^\alpha)\in \R^l$
is a vector from dilatonic scalar fields,
$(C_{\alpha\beta})$ is a non-degenerate $l\times l$ matrix ($l\in \N$),
$\theta_a = \pm 1$,
\beq{2.2}
F^a =  dA^a
\eeq
is a $n_a$-form ($n_a \geq 2$) on a $D$-dimensional manifold $M$,
$\Lambda$ is a cosmological constant
and $\lambda_{a}$ is a $1$-form on $\R^l$ :
$\lambda_{a} (\varphi) =\lambda_{a \alpha}\varphi^\alpha$,
$a \in \Delta$, $\alpha=1,\ldots,l$.
In (\ref{2.1})
we denote $|g| = |\det (g_{MN})|$,
\beq{2.3}
(F^a)^2_g =
F^a_{M_1 \ldots M_{n_a}} F^a_{N_1 \ldots N_{n_a}}
g^{M_1 N_1} \ldots g^{M_{n_a} N_{n_a}},
\eeq
$a \in \Delta$, where $\Delta$ is some finite set, and $S_{\rm GH}$ is the
standard Gibbons-Hawking boundary term \cite{GH}. In the models
with one time all $\theta_a =  1$  when the signature of the metric
is $(-1,+1, \ldots, +1)$.

The equations of motion corresponding to  (\ref{2.1}) have the following
form
\bear{2.4}
R_{MN} - \frac{1}{2} g_{MN} R  =   T_{MN} - \Lambda g_{MN},
\\
\label{2.5}
{\btu}[g] \varphi^\alpha -
\sum_{a \in \Delta} \theta_a  \frac{\lambda^{\alpha}_a}{n_a!}
e^{2 \lambda_{a}(\varphi)} (F^a)^2_g = 0,
\\
\label{2.6}
\nabla_{M_1}[g] (e^{2 \lambda_{a}(\varphi)}
F^{a, M_1 \ldots M_{n_a}})  =  0,
\ear
$a \in \Delta$; $\alpha=1,\ldots,l$.
In (\ref{2.5}) $\lambda^{\alpha}_{a} = C^{\alpha \beta}
\lambda_{\beta a}$, where $(C^{\alpha \beta})$
is matrix inverse to $(C_{\alpha \beta})$.
In (\ref{2.4})
\bear{2.7}
T_{MN} =   T_{MN}[\varphi,g]
+ \sum_{a\in\Delta} \theta_a  e^{2 \lambda_{a}(\varphi)} T_{MN}[F^a,g],
\ear
where
\bear{2.8}
T_{MN}[\varphi,g] =
C_{\alpha\beta}\left(\p_{M} \varphi^\alpha \p_{N} \varphi^\beta -
\frac{1}{2} g_{MN} \p_{P} \varphi^\alpha \p^{P} \varphi^\beta\right),
\\
T_{MN}[F^a,g] = \frac{1}{n_{a}!}  [ - \frac{1}{2} g_{MN} (F^{a})^{2}_{g}
 + n_{a}  F^{a}_{M M_2 \ldots M_{n_a}} F_{N}^{a, M_2 \ldots M_{n_a}}].
\label{2.9}
\ear
In (\ref{2.5}), (\ref{2.6}) ${\btu}[g]$ and ${\btd}[g]$
are Laplace-Beltrami and covariant derivative operators respectively
corresponding to  $g$.

Let us consider the manifold
\beq{2.10}
M = M_{0}  \times M_{1} \times \ldots \times M_{n},
\eeq
with the metric
\beq{2.11}
g= e^{2{\gamma}(x)} g^0  +
\sum_{i=1}^{n} e^{2\phi^i(x)} g^i ,
\eeq
where $g^0  = g^0 _{\mu \nu}(x) dx^{\mu} \otimes dx^{\nu}$
is an arbitrary metric with any signature on the manifold $M_{0}$
and $g^i  = g^{i}_{m_{i} n_{i}}(y_i) dy_i^{m_{i}} \otimes dy_i^{n_{i}}$
is a metric on $M_{i}$  satisfying the equation
\beq{2.13}
R_{m_{i}n_{i}}[g^i ] = \xi_{i} g^i_{m_{i}n_{i}},
\eeq
$m_{i},n_{i}=1, \ldots, d_{i}$; $\xi_{i}= {\rm const}$,
$i=1,\ldots,n$.  Thus, $(M_i, g^i )$  are Einstein spaces.
The functions $\gamma, \phi^{i} : M_0 \rightarrow {\bf R}$ are smooth.
Here we denote $d_{\nu} = {\rm dim} M_{\nu}$; $\nu = 0, \ldots, n$.
$D = \sum_{\nu = 0}^{n} d_{\nu}$.
We claim any manifold $M_{\nu}$ to be oriented and connected.
Then the volume $d_i$-form
\beq{2.14}
\tau_i  \equiv \sqrt{|g^i(y_i)|}
\ dy_i^{1} \wedge \ldots \wedge dy_i^{d_i},
\eeq
and signature parameter
\beq{2.15}
\varepsilon(i)  \equiv {\rm sign}( \det (g^i_{m_i n_i})) = \pm 1
\eeq
are correctly defined for all $i=1,\ldots,n$.

Let $\Omega = \Omega(n)$  be a set of all non-empty
subsets of $\{ 1, \ldots,n \}$.
The number of elements in $\Omega$ is $|\Omega| = 2^n - 1$.
For any $I = \{ i_1, \ldots, i_k \} \in \Omega$, $i_1 < \ldots < i_k$,
we denote
\beq{2.16a}
\tau(I) \equiv \tau_{i_1}  \wedge \ldots \wedge \tau_{i_k},
\eeq
\beq{2.20}
M_{I} \equiv M_{i_1}  \times  \ldots \times M_{i_k},
\eeq
\beq{2.19}
d(I) \equiv  \sum_{i \in I} d_i = d_{i_1} + \ldots + d_{i_k},
\eeq
where $d_i$ is both, the dimension of the oriented manifold $M_i$
and the rank of the volume form $\tau_i$.

\subsection*{Ansatz for composite electric p-branes }

Let $\Delta_e \subset \Delta$ be a non-empty subset,
and
\bear{2.e}
j_e: \Delta_e \to P_*(\Omega) \qquad \qquad \quad
\nn \\
\qquad  a \mapsto \Omega_{a,e} \in \Omega,
\quad \Omega_{a,e} \neq \emptyset
\ear
a map from $\Delta_e$ into the set $P_*(\Omega)$ of all non-empty subsets
of $\Omega$,
satisfying the condition
\beq{2.23}
d(I) + 1 = n_a ,
\eeq
for all $I \in \Omega_{a,e}$, $a \in \Delta_e$.
In the following we fix the map \rf{2.e}.

For the potential forms $A^a$, $a \in \Delta_e $, we make the ansatz
\bear{2.e1}
A^a = A^{a,e}= 0, \qquad a \in \Delta \setminus \Delta_e,
\\
\label{2.e2}
A^a = A^{a,e} =\sum_{I \in \Omega_{a,e}} A^{a,e,I}, \qquad a \in \Delta_e,
\ear
where, with $\tau(I)$ from (\ref{2.16a}),
\beq{2.17}
A^{a,e,I} = \Phi^{a,e,I} (x) \tau(I),
\eeq
are elementary, electric type potential forms,
with functions $\Phi^{a,e,I}$ smooth on $M_0$,
$I \in \Omega_{a,e}$, $a \in \Delta_e$.

It follows from \rf{2.e1}-\rf{2.17} that
\bear{2.e3}
F^a = F^{a,e}=0, \qquad a \in \Delta \setminus \Delta_e
\\
\label{2.e4}
F^a = F^{a,e}=\sum_{I \in \Omega_{a,e}} F^{a,e,I}, \qquad a \in \Delta_e,
\ear
where
\beq{2.21}
F^{a,e,I} = dA^{a,e,I} = d \Phi^{a,e,I} \wedge \tau(I).
\eeq
Due to \rf{2.23} this relation is indeed self-consistent.

For dilatonic scalar fields we put
\beq{2.24}
\varphi^\alpha = \varphi^\alpha(x),
\eeq
$\alpha=1,\ldots,l$.
Thus, in our ansatz all fields
depend just on the point $x \in M_0$.

{\bf Remark 1.} It is more correct to write in
(\ref{2.11}) $\hat{g}^{\alpha}$ instead  $g^{\alpha}$,
where $\hat{g}^{\alpha} =  p_{\alpha}^{*} g^{\alpha}$ is the
pullback of the metric $g^{\alpha}$  to the manifold  $M$
by the canonical projection: $p_{\alpha} : M \rightarrow M_{\alpha}$,
$\alpha = 0, \ldots, n$.
Analogously, we should write $\hat{\Phi}^{a,e,I}$ and $\hat{\tau}(I)$
instead  $\Phi^{a,e,I}$ and $\tau(I)$  in (\ref{2.17}) and
(\ref{2.21}). In what follows
we omit all "hats" in order to simplify notations.

\subsection*{Ricci-tensor components}

The nonzero Ricci tensor components
for the metric (\ref{2.11}) are the following \cite{IM}
\bear{2.25}
R_{\mu \nu}[g]  =   R_{\mu \nu}[g^0 ] +
          g^0 _{\mu \nu} \Bigl[- \Delta_0 \gamma
          +(2-d_0)  (\p \gamma)^2
- \p \gamma \sum_{j=1}^{n} d_j \p \phi^j ]
\\ \nn
+ (2 - d_0) (\gamma_{;\mu \nu} - \gamma_{,\mu} \gamma_{,\nu})
 - \sum_{i=1}^{n} d_i ( \phi^i_{;\mu \nu} - \phi^i_{,\mu} \gamma_{,\nu}
 - \phi^i_{,\nu} \gamma_{,\mu} + \phi^i_{,\mu} \phi^i_{,\nu}),
\\
\label{2.26}
R_{m_{i} n_{i}}[g]  = {R_{m_{i} n_{i}}}[g^i ]
     - e^{2 \phi^{i} - 2 \gamma} g^i _{m_{i} n_{i}}
      \biggl\{ \Delta_0 \phi^{i}
+ (\p \phi^{i}) [ (d_0 - 2) \p \gamma  +
          \sum_{j=1}^{n} d_j \p \phi^j ] \biggr\},
\ear
Here
$\p \beta \,\p \gamma \equiv g^{0\  \mu \nu} \beta_{, \mu} \gamma_{, \nu}$
and  $\Delta_0$ is the Laplace-Beltrami operator corresponding
to  $g^0 $. The scalar curvature for (\ref{2.11}) is \cite{IM}
\bear{2.27}
  R[g] =  \sum_{i =1}^{n} e^{-2 \phi^i} {R}[g^i ]
          + e^{-2 \gamma} \biggl\{ {R}[g^0 ]
          - \sum_{i =1}^{n} d_i (\p \phi^i)^2
\\ \nn
  -  (d_0 {-} 2) (\p \gamma)^2
     -  (\p f)^2 - 2 \Delta_0 (f +  \gamma) \biggr\},
\ear
where
\beq{2.28}
f = {f}(\gamma, \phi)  = (d_0 - 2) \gamma +\sum_{j=1}^{n} d_j  \phi^j .
\eeq

\subsection*{$\sigma$-model representation}

{\bf Restriction 1.} Let us start first from the case
\beq{2.r}
n_1 \equiv | \{ i \mid d_i=1,i\geq 1\} | \leq 1,
\eeq
i.e. the number of $1$-dimensional manifolds
$M_i$, $i>0$, in \rf{2.10} is not more than $1$.
In this case the energy momentum tensor  $T_{MN}$ from \rf{2.7}
has a block diagonal form (see Subsect. 4.2 below) that
assures the existence of a $\sigma$-model representation.
In Subsection 4.2 the restriction   (\ref{2.r}) will be
omitted.

Using (\ref{2.25}), (\ref{2.26}) and (\ref{2.r}),
it is not difficult to verify
(see Proposition 2 and  Remark 2 in Sect. 6 below)
that the field equations
(\ref{2.4})-(\ref{2.6}) for the field configurations
>from  (\ref{2.11}), (\ref{2.e3})-(\ref{2.21}) and (\ref{2.24})
may be obtained as the equations of motion from the action
\bear{2.29}
S_{\sigma} = S_{\sigma}[g^0 , \gamma,\phi,\varphi,\Phi]
 =   \frac{1}{2 \kappa^{2}_0}
     \int_{M_0} d^{d_0}x   \sq {g^{0}}
     e^{f(\gamma, \phi)} \biggl\{ {R}[g^0 ]
- \sum_{i =1}^{n} d_i (\p \phi^i)^2
\\ \nn
-  (d_0 - 2) (\p \gamma)^2   + (\p f) \p (f + 2\gamma)
+      \sum_{i=1}^{n} \xi_{i} d_i e^{-2 \phi^i + 2 \gamma} -
     2 \Lambda e^{2 \gamma} - {\cal L}  \biggr\},
\ear
where
\bear{2.30a}
{\cal L} = {\cal L}_{\varphi} + {\cal L}_A,
\\
\label{2.30b}
{\cal L}_{\varphi}
= C_{\alpha\beta} \p \varphi^\alpha \p \varphi^\beta,
\\
\label{2.30}
{\cal L}_{A} = {\cal L}_{A,e} =
\sum_{a \in \Delta_e} \theta_a \sum_{I \in \Omega_{a,e}}
\varepsilon(I)
\exp(2 \lambda_a(\varphi) - 2 \sum_{i \in I} d_i \phi^i)
(\p \Phi^{a,e,I})^2 ,
\ear
where $|g^{0}|= |\det (g^0_{\mu\nu})|$ and similar notations are applied
to the metrics $g^{i}$, $i=1, \ldots, n$.
In (\ref{2.30})
\beq{2.31}
\varepsilon(I) \equiv
\varepsilon(i_1) \times \ldots \times \varepsilon(i_k) = \pm 1
\eeq
for $I = \{i_1, \ldots, i_k \} \in \Omega$,
$i_1 < \ldots < i_k$ (see (\ref{2.15})).

For finite internal space volumes (e.g. compact $M_i$)
$V_i$ the action (\ref{2.29}) (with ${\cal L}$ from (\ref{2.30a}))
coincides with the action (\ref{2.1}), i.e.
\beq{2.33}
  S_{\sigma}[g^0 , \gamma,\phi,\varphi,\Phi] =
  S[g(g^0,\gamma,\phi), \varphi, F(\Phi)],
\eeq
where $g = g(g^0,\gamma,\phi)$  and $F = F(\Phi)$
are defined by the relations (\ref{2.11}) and
(\ref{2.e3})-(\ref{2.21}) respectively and
\beq{2.34}
\kappa^{2} = \kappa^{2}_0 \prod_{i=1}^{n} V_i.
\eeq
This may be readily verified using the scalar curvature decomposition
\bear{2.35}
R[g] =  \sum_{i=1}^{n} \! e^{-2 \phi^i} {R}[g^i ]
          + e^{-2 \gamma} \Bigl\{ {R}[g^0 ]
          - \sum_{i =1}^{n} d_i (\p \phi^i)^2
\\ \nn
- (d_0 - 2) (\p \gamma)^2 + (\p f) \p (f + 2 \gamma) + R_{B} \Bigr\},
\ear
where
\bear{2.36}
R_B = (1/\sq {g^0 }) e^{-f}
     \p_{\mu} [-2 e^f \sq {g^0 }
     g^{0\ \mu \nu} \p_{\nu} (f + \gamma)]
\ear
gives rise to the Gibbons-Hawking boundary term
\beq{2.37}
S_{\rm GH} = \frac{1}{2\kappa^{2}} \int_{M} d^{D}z \sq g
     \{ - e^{-2 \gamma} R_{B} \}.
\eeq
We note that (\ref{2.30}) appears
due to the relation
\beq{2.38}
\frac{1}{n_a!} (F^{a,e,I})^2
= \varepsilon(I)
\exp(- 2 \gamma  - 2 \sum_{i \in I} d_i \phi^i) (\p \Phi^{a,e,I})^2 ,
\eeq
$I \in \Omega_{a,e}$, $a \in \Delta_e$.

\section{Exact solutions}
\setcounter{equation}{0}

First we consider the case $d_0 \neq 2$.
In order to simplify the action (\ref{2.29}), we use,
as in ref. \cite{IM4,IM},  for $d_0 \neq 2$ the
generalized  harmonic gauge
\beq{3.1}
\gamma = {\gamma}_{0}(\phi) =
\frac{1}{2- d_0}  \sum_{i =1}^{n} d_i \phi^i.
\eeq
It means that $f = {f}(\gamma_0, \phi)= 0$.
This gauge does not exist for $d_0 = 2$.
For the cosmological case  with $d_0 =1$ and  $g^0  = - dt \otimes dt$,
the gauge (\ref{3.1}) is the harmonic-time gauge
\cite{IM1,IMZ} (for spherical symmetry see K.A.Bronnikov, 1973 \cite{Br}).

>From equations (\ref{2.29}), (\ref{2.30}), (\ref{3.1}) we get
\bear{3.2}
S_{0}[g^0 ,\phi, \varphi,\Phi] =
S_{\sigma}[g^0,\gamma_0(\phi),\phi,\varphi,\Phi] =
\frac{1}{2 \kappa^{2}_0}
     \int_{M_0} d^{d_0}x \sq {g^0 }
\Bigl\{
{R}[g^0]
\\ \nn
- G_{ij} g^{0\ \mu \nu} \p_{\mu} \phi^i  \p_{\nu} \phi^j -
2 {V}(\phi) - {\cal L}
\Bigr\},
\ear
where
\beq{3.3}
G_{ij} = d_i \delta_{ij} + \frac{d_i d_j}{d_0 -2}
\eeq
are the components of the ("purely gravitational")
midisuperspace  metric on
${\bf R}^{n}$  \cite{IM}
(or the gravitational part of target space metric),
$i, j = 1, \ldots, n$,
and
\beq{3.4}
 V = {V}(\phi)
= \Lambda e^{2 {\gamma_0}(\phi)}
-\frac{1}{2}   \sum_{i =1}^{n} \xi_i d_i e^{-2 \phi^i
+ 2 {\gamma_0}(\phi)}
\eeq
is the potential and ${\cal L}$ defined in (\ref{2.30a}).
This is the action of a self-gravitating
$\sigma$ model on  $M_0$ with a
$(n + l + \sum_{a\in \Delta_e}| \Omega_{a,e}|)$-dimensional target space
and a self-interaction described by the
potential (\ref{3.4}).

\subsection{$\sigma$-model with zero potential.}

Now we consider the case $\xi_i = \Lambda = 0$,
i.e. all spaces $(M_i, g^i)$ are Ricci-flat,
$i = 1, \ldots, n$, and the cosmological constant is zero.
In this case the potential  (\ref{3.4}) is trivial
and we are led to the $\sigma$-model with the
action
\bear{3.5}
S_{\sigma} =  \int_{M_0} d^{d_0}x \sq {g^0 } \Bigl\{ {R}[g^0]
- \hat{G}_{AB} \p \sigma^A \p \sigma^B
- \sum_{s \in S} \varepsilon_s e^{2 L_{A s} \sigma^A}
(\p \Phi^s)^2  \Bigr\},
\ear
where we put $2 \kappa^{2}_0 =1$.
In (\ref{3.5})
$(\sigma^A) = (\phi^i, \varphi^\alpha) \in {\bf R}^{N}$,
where $N = n + l$,
\beq{3.6}
 \left(\hat{G}_{AB} \right)    =
                             \left(
                              \begin{array}{cc}
                                G_{ij} &  0 \\
                                   0   &  C_{\alpha \beta}
                               \end{array}
                          \right)
\eeq
is a non-degenerate (block-diagonal) $N \times N$-matrix,
\beq{3.s}
S = S_e \equiv \bigsqcup_{a \in \Delta_e} \{a\}
\times \{ e \} \times\Omega_{a,e},
\eeq
and for $s=(a,e,I)\in S_e$; $a\in \Delta_e$;  $I \in \Omega_{a,e}$
we denote
\beq{3.e}
\eps_s= \theta_a \eps(I)=\pm 1,
\eeq
$\Phi^s=\Phi^{a,e,I}$ and  vectors
\beq{3.7}
L_s = (L_{As}) = (L_{i s}, L_{\alpha s})
= (l_{i I}, \lambda_{\alpha a}) \in {\bf R}^N
\eeq
$s \in S_e$, are defined by relation
\beq{3.8}
l_I = (l_{jI}) \equiv (- \sum_{i \in I} d_i \delta^i_j)  \in {\bf R}^n,
\eeq
$i,j = 1, \ldots, n$; $\alpha =1,\ldots,l$; $a \in \Delta_e$.

The equations of motion corresponding to \rf{3.5} are
\bear{3.9}
R_{\mu \nu}[g^0]  =
\hat{G}_{AB} \p_{\mu} \sigma^A \p_{\nu} \sigma^B
+ \sum_{s \in S} \eps_s e^{2 L_{As} \sigma^A}
\p_{\mu} \Phi^s  \p_{\nu} \Phi^s,
\\
\label{3.10}
\hat{G}_{AB} {\btu}[g^0] \sigma^B
-  \sum_{s \in S} \eps_s L_{As} e^{2 L_{Bs} \sigma^B} (\p \Phi^s)^2 =
0, \\ \label{3.11}
\p_{\mu} \left( \sqrt{|g^0|} g^{0 \mu \nu} e^{2 L_{As}
\sigma^A} \p_{\nu} \Phi^s \right) = 0, \ear $A = i,\alpha$; $i=1, \ldots,
n$; $\alpha=1, \ldots,l$; $s \in S = S_e$.

In what follows we define
a non-degenerate (real-valued) quadratic form
\beq{3.21}
(X,Y)_* \equiv X_A \hat{G}^{AB} X_{B},
\eeq
where $(\hat{G}^{AB})= (\hat{G}_{AB})^{-1}$.

{\bf Proposition 1.} Let  $S_{*} \subset S$ be a non-empty
set of indices such that there exists a set of real non-zero
numbers $\nu_s, s \in S_{*},$  satisfying the relations
\beq{3.12}
 (L_{s}, L_{r})_* = - \eps_s (\nu_s)^{-2}\delta_{sr},
\eeq
$s,r \in S_{*}$. Let $(M_0,g^0)$ be Ricci-flat
\beq{3.13}
R_{\mu \nu}[g^0]  = 0.  \qquad  \qquad
\eeq
Then the field configuration
\bear{3.14}
\sigma^A = \sum_{s \in S_{*}} \alpha^A_s \ln H_s,
\quad
\\
\label{3.15}
\Phi^s  = \frac{\nu_s}{H_s},
\qquad
s \in S_{*},
\\
\label{3.16}
\Phi^{{s'}} = C_{s'}\in\R,
\qquad
{s'} \in  S \setminus  S_{*}
\ear
satisfies the field equations (\ref{3.9})-(\ref{3.11}) if
\beq{3.17}
\alpha^A_s = -  \hat{G}^{AB} L_{Bs} \eps_s (\nu_s)^{2},
\eeq
$A = 1, \ldots, N$; $s \in S_{*}$;
$\nu_s$ satisfy (\ref{3.12}), and functions $H_s = H_{s}(x) > 0$ are
harmonic, i.e.
\beq{3.18}
{\btu}[g^0] H_s = 0,
\eeq
$s \in S_{*}$.

Proposition 1 follows just from
substitution of (\ref{3.12})-(\ref{3.18}) into
the  equations of motion (\ref{3.9})-(\ref{3.11}).

Thus, due to (\ref{3.12}), the
vectors $L_s$, $s \in S_{*}$, are orthogonal to each other,
and $(L_{s}, L_{s})_*$ has a sign opposite to that of
$\eps_s$, $s \in S_{*}$. When the form $(\cdot,\cdot)_*$ is
positive-definite
(this take place for $d_0 > 2$ and a positive-definite
matrix $C=(C_{\alpha\beta})$,
the sign is $\eps_s = -1$ for all $s \in S_{*}$.

Now, we apply Proposition 1 to the present
model with Ricci-flat spaces  $(M_i, g^i)$, $i = 1, \ldots, n$,
and zero cosmological constant.  From (\ref{3.6}), (\ref{3.7}) and
(\ref{3.21}) we get
\beq{3.22}
(L_{s}, L_{r})_* =  \avers{l_I, l_J }  + {\lambda}_a \cdot {\lambda}_b,
\eeq
with $s=(a,e,I)$ and $r=(b,e,J)$ in $S_e$
($a,b\in\Delta_e$; $I \in \Omega_{a,e}$;  $J \in \Omega_{b,e}$).
Here $l_I$ are defined in  (\ref{3.8})  and
\beq{3.24}
{\lambda}_a \cdot {\lambda}_b
\equiv  C^{\alpha\beta} {\lambda}_{\alpha a}  {\lambda}_{\beta b},
\eeq
for
$a, b \in \Delta_e$.
In (\ref{3.22})
\beq{3.25}
\avers{u,v} \ \equiv    u_i G^{ij} v_j
\eeq
is a quadratic form on  ${\bf R}^n$. Here,
\beq{3.26}
G^{ij} =
\frac{\delta_{ij}}{d_i} + \frac{1}{2 - D}
\eeq
are components of the matrix inverse to the matrix  $(G_{ij})$ in
(\ref{3.3}).

>From (\ref{3.8}), (\ref{3.25}) and (\ref{3.26})
we obtain
\beq{3.27}
\avers{l_I, l_J} = d(I \cap J) + \frac{d(I) d(J)}{2-D},
\eeq
$I, J \in \Omega$.
In (\ref{3.27}) $d(\emptyset) = 0$.
Without restriction let $S_*=S_e$ (if initially $S_*\neq S_e$,
one may redefine $\Delta_e$ and $j_e$ from \rf{2.e} such that
$S_*=S_e$ thereafter).

Due to (\ref{3.22}) and (\ref{3.27})  the
relation (\ref{3.12}) reads
\beq{3.28}
d(I \cap J) + \frac{d(I) d(J)}{2-D}
+ C^{\alpha\beta} {\lambda}_{\alpha a}  {\lambda}_{\beta b}
= - \theta_a \eps(I) (\nu_{a,e,I})^{-2}\delta_{ab}\delta_{IJ} ,
\eeq
with $I \in \Omega_{a,e}$,
$J \in \Omega_{b,e}$, $a, b \in \Delta_{e}$,
denoting $\nu_{a,e,I} \equiv \nu_{(a,e,I)}$.

For coefficients $\alpha^A_s$ from (\ref{3.17}) we get,
for $s=(a,e,I) \in S_e=S_*$,
\bear{3.29}
\alpha^i_s
= - \theta_a G^{ij}l_{jI} \eps(I)  \nu_{a,e,I}^{2}
=  \biggl( \sum_{j \in I} \delta^i_j  + \frac{d(I)}{2-D} \biggr)
\theta_a \eps(I) \nu_{a,e,I}^{2} ,
\\\label{3.30}
\alpha^\beta_I =
- C^{\beta\gamma}\lambda_{\gamma a} \theta_a \eps(I) \nu_{a,e,I}^{2},
\ear
$i = 1, \ldots, n$; $\beta,\gamma=1,\ldots,l$.

Relations (\ref{3.14})
with $(\sigma^A) = (\phi^i, \varphi^\beta)$, $S_*=S_e$
read
\bear{3.31}
\phi^i = \sum_{s \in S_{e}} \alpha^i_s   \ln H_s ,
\\ \label{3.32}
\varphi^\beta = \sum_{s \in S_{e}} \alpha^\beta_s  \ln H_s ,
\ear
$i = 1, \ldots, n$; $\beta=1,\ldots,l$.
These relations
imply for $\gamma$ from  (\ref{3.1})
\beq{3.33}
\gamma = \sum_{s \in S_{e}}
\alpha^0_s  \ln H_s ,
\eeq where
\beq{3.34}
\alpha^0_s  = \frac{d(I)}{2-D} \theta_a \varepsilon(I) \nu_{a,e,I}^{2} ,
\eeq
for $s=(a,e,I) \in S_e=S_*$,
$a \in \Delta_e; I \in \Omega_{a,e}$.

\subsection*{The solution.}

Thus, the equations of motion (\ref{2.4})-(\ref{2.6}) with  $\Lambda = 0$
defined on the manifold (\ref{2.10}) have the following solution:
\bear{4.1}
g=U_e \{g^0+\sum_{i=1}^n U_{i,e} g^i\}, \\
\label{4.1e}
U_e\equiv
\left(\prod_{a\in\tri_e} \prod_{I\in\Omega_{a,e}}
H_{a,e,I}^{2\theta_a\eps(I)d(I)
\nu_{a,e,I}^2}\right)^{\frac1{2-D}}, \\
\label{5.1i}
U_{i,e}\equiv
\prod_{a\in\tri_e} \prod_{I\in\Omega_{a,e}, \ I \ni i}
H_{a,e,I}^{2\theta_a\eps(I)\nu_{a,e,I}^2}, \\
\label{4.p}
\varphi^{\beta} = \varphi^{\beta}_e \equiv
- \sum_{a \in \Delta_e} \theta_a \sum_{I \in \Omega_{a,e}}
C^{\beta\gamma}\lambda_{\gamma a} \eps(I)
\nu^2_{a,e,I} \ln H_{a,e,I},
\\ \label{4.a1}
F^{a} = F^{a,e} =
 \sum_{I \in \Omega_{a,e}}
\nu_{a,e,I} d H^{-1}_{a,e,I} \wedge \tau(I) ,
\qquad a \in \Delta_e ,
\\
\label{4.a2}
F^{a} = 0 ,
\qquad a \not \in \Delta_e
\ear
(we put $\prod_{\emptyset} \ldots \equiv 1$ ),
where $\beta=1,\ldots,l$; $a \in \Delta_e$;
forms $\tau(I)$ are defined in \rf{2.16a},
parameters $\nu_s \neq 0$ and ${\lambda}_a$
satisfy the relation (\ref{3.28}),
and  functions
$H_s = H_{s}(x) > 0$,
are harmonic on $(M_0, g^0)$, i.e.
\beq{3.h}
{\btu}[g^0] H_s = 0 ,
\eeq
$s \in S_{e}$.
In (\ref{4.1})
\beq{3.36}
{\rm Ric}[g^0]= {\rm Ric}[g^1] = \ldots = {\rm Ric}[g^n] = 0
\eeq
(${\rm Ric}[g^{\nu}]$ is Ricci-tensor corresponding to $g^{\nu}$).
Relations (\ref{3.28}) read
\bear{4.2}
- \eps(I) \theta_a (\nu_{a,e,I})^{-2} =
d(I) + \frac{(d(I))^2}{2-D}
+ C^{\alpha\beta} {\lambda}_{\alpha a}  {\lambda}_{\beta b},
\qquad  \\
\label{4.3}
d(I \cap J) + \frac{d(I) d(J)}{2-D} +
 C^{\alpha\beta} {\lambda}_{\alpha a}  {\lambda}_{\beta b} = 0 ,
\qquad (a,I) \neq (b,J) ,
\ear
where $I \in \Omega_{a,e}$; $J \in \Omega_{b,e}$;
$a,b \in \Delta_e$.

The solution presented here is valid also for $d_0 = 2$.
It may be verified using the $\sigma$-model
representation (\ref{2.29}) for $d_0 = 2$
with $f=0$.

Note that, for positive definite matrix $(C_{\alpha\beta})$
(or $(C^{\alpha \beta})$) and $d_0 \geq 2$, \rf{3.28} implies
(cf. \cite{IM4}, Proposition 2)
\beq{3.43}
\eps(I)=-\theta_a ,
\eeq
for all $I \in \Omega_{a,e}$; $a \in \Delta_e$.
Therefore, for $\theta_a = 1$
the restriction $g_{\vert M_{I}}$ of the metric \rf{2.11}
to a membrane manifold $M_{I}$ has an odd number
of linearly independent timelike directions.

However, if the metric $(C_{\alpha\beta})$ in the space of scalar fields
is not positive-definite
 (this takes place for $D = 12$  model from \cite{KKP}),
then \rf{3.43}   may be violated for sufficently negative
$\lambda_a^2 =
C^{\alpha\beta} {\lambda}_{\alpha a}  {\lambda}_{\beta b}<0$.
In this case a non-trivial potential $A^{a}$ may also exist
on an Euclidean p-brane for $\theta_a =1$.

\section{Calculations for energy-momentum tensor and
additional constraints}
In this section we omit the restriction \rf{2.r} and show that in
general case some additional constraints should be imposed on the
$\sigma$-model \rf{2.29}.

\subsection{Useful relations}
Let $F_1$ and $F_2$ be forms of rank $r$ on $(M,g)$ ($M$ is a manifold and
$g$ is a metric on it). We define
\bear{4.1n}
(F_1\cdot F_2)_{MN}\equiv
{(F_1)_{MM_2\dots M_r}(F_2)_N}^{M_2\cdots M_r};
\\ \label{4.2n}
F_1F_2 \equiv {(F_1\cdot F_2)_M}^M=
(F_1)_{M_1\dots M_r}(F_2)^{M_1\dots M_r}.
\ear
Clearly that
\beq{4.3n}
(F_1\cdot F_2)_{MN}=(F_2\cdot F_1)_{NM},\quad F_1F_2=F_2F_1.
\eeq

For the form $F^{a,e,I}$ from \rf{2.21} and metric $g$ from \rf{2.11}
we obtain
\bear{4.4}
\frac1{n_a!}(F^{a,e,I}\cdot F^{a,e,I})_{\mu\nu}=
\frac{A(I)}{n_a}
\partial_\mu\Phi^{a,e,I}\partial_\nu\Phi^{a,e,I} \exp(2\gamma);
\\ \label{4.5}
\frac1{n_a!}(F^{a,e,I}\cdot F^{a,e,I})_{m_in_i}=
g^i_{m_in_i} \frac{A(I)}{n_a}(\partial\Phi^{a,e,I})^2
\exp(2\phi^i),
\ear
where $i\in I$, indices $m_i, n_i$ correspond to the manifold $M_i$ and
\beq{4.6}
A(I)\equiv A(I,\gamma,\phi)=
\eps(I)\exp\Bigl(-2\gamma-2\sum_{i\in I}d_i\phi^i \Bigr),
\eeq
$I\in\Omega_{a,e}$. All other components of $(F^{a,e,I}\cdot
F^{a,e,I})_{MN}$ are zero. For the scalar invariant we have
\beq{4.7}
\frac1{n_a!}(F^{a,e,I})^2\equiv\frac1{n_a!}F^{a,e,I}F^{a,e,I}=
A(I)(\partial\Phi^{a,e,I})^2,
\eeq
$I\in\Omega_{a,e}$. We recall that here, as above, we use the notations:
$\partial\Phi_1\partial\Phi_2=
g^{0\mu\nu}\partial_\mu\Phi_1\partial_\nu\Phi_2$,
$(\partial\Phi_1)^2=\partial\Phi_1\partial\Phi_1$ for functions
$\Phi_1=\Phi_1(x)$, $\Phi_2=\Phi_2(x)$ on $M_0$.

Now  consider the tensor field
\beq{4.8}
(F^{a,e,I}\cdot F^{a,e,J})_{MN}dz^M\otimes dz^N
\eeq
for $I\ne J$; $I,J\in\Omega_{a,e}$. From \rf{2.23} we get $d(I)=d(J)$ and
hence
\beq{4.9}
I\ne I\cap J; \quad J\ne I\cap J.
\eeq
Indeed, if we suppose, for example, that $I\cap J=I$, then we obtain
(see \rf{2.19})
\beq{4.10}
d(J)=d(I\cap J)+d(J\setminus I)=d(I)+d(J\setminus I)
\eeq
or, equivalently, $d(J\setminus I)=0\Iff J=I$. But $I\ne J$.

It may be easily verified that for $I\ne J$ the scalar invariant is
trivial:
\beq{4.11}
F^{a,e,I}F^{a,e,J}=0.
\eeq

Now we present the non-zero components for the tensor \rf{4.8}. Let
\beq{4.12}
w_1\equiv\{i \mid i\in\{1,\dots,n\},\ d_i=1\}.
\eeq
The set $w_1$ describes all $1$-dimensional manifolds among $M_i$ $(i\ge1)$.

It may be verified by a straight-forward calculation that the tensor
\rf{4.8} may be non-zero  only if
\beq{4.13}
n_1 = |w_1| \ge 2,
\eeq
i.e. the number of one-dimensional manifolds among $M_i$, $i\ge1$, is more
than 1. The only possible non-zero components of \rf{4.8} for $I\ne J$;
$I,J\in\Omega_{a,e}$, are the following
\bear{4.14}
\frac1{(n_a-1)!}(F^{a,e,I}\cdot F^{a,e,J})_{1_i1_j}=
\delta(i,I\cap J)\delta(j,I\cap J)\eps(I\cap J)
\sqrt{|g^i|\ |g^j|} \nn \\
\times\exp\Bigl(-2\gamma-2\sum_{l\in I\cap J}d_l\phi^l\Bigr)
\partial\Phi^{a,e,I}\partial\Phi^{a,e,J},
\ear
where $i\ne j$; $i,j\in w_1$; $\delta(i,{\cal K})=\pm1$ is defined for
$\{i\}\sqcup{\cal K}\in\Omega_{a,e}$ $(i\notin{\cal K})$ by the relation
\beq{4.15}
\delta(i,{\cal K})\tau(\{i\}\sqcup{\cal K})=\tau_i\wedge\tau({\cal K}).
\eeq
(The volume form $\tau(I)$ is defined in \rf{2.16a}.) We put
$\tau(\emptyset)= \delta(i,\emptyset)=\eps(\emptyset)=1$.
Here and in what follows we denote
\beq{4.15d}
(A \sqcup B = C)
\Iff (A \cup B = C, A \cap B = \emptyset).
\eeq

\subsection{Energy-momentum tensor and constraints}
For the "composite" field $F^{a,e}$, $a\in\tri_e$, from \rf{2.e4} we
have
\beq{4.16}
(F^{a,e})^2=\Bigl(\sum_{I\in\Omega_{a,e}}F^{a,e,I}\Bigr)^2=
\sum_{I\in\Omega_{a,e}}(F^{a,e,I})^2
\eeq
and
\beq{4.17}
(F^{a,e}\cdot F^{a,e})_{MN}=
\sum_{I\in\Omega_{a,e}}(F^{a,e,I}\cdot F^{a,e,I})_{MN}+
\sum_{I,J\in\Omega_{a,e}\atop I\ne J}(F^{a,e,I}\cdot F^{a,e,J})_{MN}.
\eeq

The relations \rf{4.16} and \rf{4.17} imply the following relations for
the energy-momentum tensor corresponding to $F^{a,e}$ (see \rf{2.9})
\beq{4.18}
T_{MN}[F^{a,e},g]=\sum_{I\in\Omega_{a,e}}T_{MN}[F^{a,e,I},g]+
\bar T_{MN}[F^{a,e},g],
\eeq
where
\beq{4.19}
\bar T_{MN}[F^{a,e},g]\equiv\frac1{(n_a-1)!}
\sum_{I,J\in\Omega_{a,e}\atop I\ne J}(F^{a,e,I}\cdot F^{a,e,J})_{MN}.
\eeq

Using the results from the previous subsection we obtain that the non-zero
components for $\bar T_{MN}$ may take place only if the condition \rf{4.13}
holds and in this case
\bear{4.20}
T_{1_i1_j}[F^{a,e},g]=\bar T_{1_i1_j}[F^{a,e},g] \nn \\
=\sqrt{|g^i|}\sqrt{|g^j|}\exp(-2\gamma)C_{ij}(\Phi^{a,e},\varphi,\phi,g^0),
\ear
where
\bear{4.21}
C_{ij}= C_{ij}(\Phi^{a,o},\phi,g^0)\equiv
\sum_{(I,J)\in W_{ij}(\Omega_{a,o})}\delta(i,I\cap J)
\delta(j,I\cap J)\eps(I\cap J) \nn \\
\times\exp\Bigl(-2\sum_{l\in I\cap J}d_l\phi^l\Bigr)
g^{0\mu\nu}\partial_\mu\Phi^{a,o,I}\partial_\nu\Phi^{a,o,J},
\ear
$a\in \tri_o$; $i,j\in w_1$; $i\ne j$ $(|w_1|\ge2)$
$\Phi^{a,o} = (\Phi^{a,o,I},I\in\Omega_{a,o})$. Here
$o = e$ and
\bear{4.22}
W_{ij}(\Omega_{1})\equiv
\{(I,J)|I,J\in\Omega_{1},\ I=\{i\}\sqcup(I\cap J),\
J=\{j\}\sqcup(I\cap J)\}.
\ear
$i,j\in w_1$, $i\ne j$, $\Omega_{1} \subset \Omega$.

The non-block-diagonal part of the total energy-momentum tensor \rf{2.7}
has the following form
\beq{4.25}
T_{1_i1_j}[F^e,\varphi,g]=\sqrt{|g^i|}\sqrt{|g^j|}
\e^{-2\gamma}C_{ij}^e(\Phi^e, \varphi, \phi,g^0),
\eeq
where  $\Phi^{e} = (\Phi^{a,e})$ and
\bear{4.26}
C_{ij}^e(\Phi^{e},\varphi,\phi,g^0)\equiv
\sum_{a\in\tri_e}\theta_a\exp[2\lambda_a(\varphi)]
C_{ij}(\Phi^{a,e},\phi,g^0),
\ear
$i,j\in w_1$; $i\ne j$. Relations \rf{4.25}, \rf{4.26} follow from
relations \rf{2.7}, \rf{4.20} and \rf{4.21}.

>From Einstein-Hilbert equations \rf{2.4} $(\Lambda=0)$ and
block-diagonal form of metric and Ricci-tensor (in the considered ansatz)
it follows that
\beq{4.27}
T_{1_i1_j}=0,
\eeq
$i \neq j$, and hence
\beq{4.28}
C_{ij}^e(\Phi^e,\varphi,\phi,g^0)=0, \quad i<j,
\eeq
$i,j\in w_1$ ($C_{ij}^e = C_{ji}^e$).

Thus, we obtain
\beq{4.29}
m_1=n_1(n_1-1)/2
\eeq
constraints, where $n_1=|w_1|\ge2$ is the number of $1$-dimensional
manifolds among $M_i$ $(i\ge1)$. The equation of motions \rf{2.5},
\rf{2.6} and block-diagonal part of \rf{2.4} are equivalent to the
equations of motion for the $\sigma$-model \rf{2.29} (or \rf{3.2}
when the harmonic gauge is fixed). The non-block-diagonal part of
\rf{2.4} leads to $m_1$ constraints on the fields of $\sigma$-model
\rf{2.29} (or \rf{3.2}).

{\bf Restriction 2e.} It follows from the presented above consideration
that the constraints are absent or are identically
satisfied in the following two cases:
i) $n_1\le1$ \rf{2.r}; ii) $n_1>1$ and
\beq{4.30}
W_{ij}(\Omega_{a,e})= \emptyset,
\eeq
(see (\ref{4.22}))
$i<j$; $i,j\in w_1$, $a\in\tri_e$. The condition \rf{4.30} means that
for any $a\in\tri_e$ and  $i<j$
($i,j=1,\dots,n$) such that $d_i=d_j=1$, there
are no sets $I,J\in\Omega_{a,e}$, such that
$I=\{i\}\sqcup(I\cap J)$ and $J=\{j\}\sqcup(I\cap J)$.
Literally (or physically) speaking the $p$-branes "feel" the presence
of additional (internal)  $1$-dimensional
directions (for example, times). The
non-trivial constraints in the world with several times occur, when there
are at least two $p$-branes
(with the same $p$) "living" in different "times"
and charged by the same field of form $F^a$ $(a\in\tri_e)$.

Thus, we weakened the Restriction 1 \rf{2.r} by adding the additional
restrictions on $p$-branes \rf{4.30}. The exact solutions of
Sect. 3 also take place  for the more general case satisfying
Restriction 2e.

\section{Magnetic p-branes}

Here we consider the ansatz for magnetically charged $p$-branes that is
dual to the one considered in subsection 2.1.

\subsection{Ansatz for composite magnetic $p$-branes}

Let $\tri_m\subset\tri$ be a non-empty subset and
\bear{5.1}
j_m: \Delta_m \to P_*(\Omega) \qquad \qquad \quad \nn \\
\qquad  a \mapsto \Omega_{a,m} \in \Omega,
\quad \Omega_{a,m} \neq \emptyset
\ear
is a map satisfying the condition
\beq{5.2}
n_a=D-d(I)-1,
\eeq
for all $I\in\Omega_{am}$; $a\in\tri_m$.

For the potential forms $A^a$, $a\in\tri_m$, we make the ansatz
\bear{5.3}
A^a = A^{a,m}=
0, \qquad a \in \Delta \setminus \Delta_m, \\
\label{5.4}
A^a = A^{a,m}=
\sum_{I \in \Omega_{a,m}} A^{a,m,I}, \qquad a \in \Delta_m,
\ear
where
\beq{5.7}
F^{a,m,I} \equiv dA^{a,m,I} =
\e^{-2\lambda_a(\varphi)}*(d\Phi^{a,m,I}\wedge \tau(I)),
\eeq
and $\Phi^{a,m,I}$: $M_0\to{\bf R}$ are smooth functions on $M_0$ and
volume forms $\tau(I)$ are defined in \rf{2.16a}, $I\in\Omega_{a,m}$;
$a\in\tri_m$. In \rf{5.7} $*=*[g]$ is Hodge operator on $(M,g)$.

>From \rf{5.7} Bianchi identities follow
\beq{5.9}
dF^{a,m,I}=0,
\eeq
$a\in\tri_m$; $I\in\Omega_{a,m}$.

In general case relations \rf{5.9} guarantee (at least) the local
existence of forms $A^{a,m,I}$ satisfying \rf{5.7}.

The field equations \rf{2.6} corresponding to $A^{a,m}$
(generalized Maxwell equations) written in the equivalent form
\beq{5.10}
*d*(\e^{2\lambda_a(\varphi)}F^{a,m})=0
\eeq
are satisfied identically for the ansatz \rf{5.3}--\rf{5.7}.

Now we impose, as in Section 2, Restriction 1. Then it may
be verified that field equations \rf{2.4}, \rf{2.5} and Bianchi relations
\rf{5.9} may be obtained as the equations of motion for the $\sigma$-model
with the action \rf{2.29} and
\beq{5.11}
{\cal L}_A= {\cal L}_{A,m}\equiv
-\eps_g\sum_{a\in\tri_m}\theta_a\sum_{I\in\Omega_{a,m}}
\eps(I)
\exp(-2\lambda_a(\varphi)-2\sum_{i\in I}d_i\phi^i)
(\partial\Phi^{a,m,I})^2,
\eeq
where $\eps_g \equiv \sign\det(g_{MN})$.

It should be noted that the sign in  \rf{5.11} is
opposite to that obtained by a straightforward substitution of
\rf{5.3}--\rf{5.7} into action \rf{2.1}. This follows from the relation
\rf{2.38} and the following identity
\beq{5.12}
\frac1{k!}F^2=\frac{\eps_g}{k_*!}(*F)^2,
\eeq
where $k=\rank F$, $k_*=\rank(*F)=D-k$. The reason for the appearance of the
additional sign "-" in \rf{5.11} can be easily explained by using the
following relation for energy-momentum tensor
\beq{5.13}
T_{MN}[*F,g]=-\eps_gT_{MN}[F,g].
\eeq
Relations \rf{5.12} and \rf{5.13} follow from the formulas
\bear{5.14}
\frac1{k_*!}(*F_1)(*F_2)=\frac{\eps_g}{k!}F_1F_2; \\
\label{5.15}
\frac1{(k_*-1)!}[(*F_1)\cdot(*F_2)]_{MN}=
\frac{\eps_g}{k!}\{g_{MN}(F_1F_2)-k(F_2\cdot F_1)_{MN}\},
\ear
where $k=\rank F_i$, $k_*=\rank(*F_i)$, $i =1,2$.
\subsection{Exact solutions}

For $d_0\ne2$, $\xi_i=\Lambda=0$, $i=1,\dots,n$,
as in pure electric case we are led to the
$\sigma$-model \rf{3.5} (we  consider
the harmonic gauge \rf{3.1}) with the set
\beq{5.16}
S=S_m  \equiv
\bigsqcup_{a \in \Delta_m}
\{a\}\times\{m\}\times\Omega_{a,m},
\eeq
and for $s=(a,m,I)\in S_m$, $I\in \Omega_{a,m}$, $a \in \tri_m$
\beq{5.17}
\eps_s=-\eps_g\theta_a\eps(I)=\pm1,
\eeq
$\Phi^s=\Phi^{a,m,I}$ and
\beq{5.18}
L_s = (L_{As}) = (L_{i s}, L_{\alpha s})
= (l_{i I}, - \lambda_{\alpha a}) \in {\bf R}^N,
\eeq
$s \in S_m$, with $l_{jI}$ defined in \rf{3.8}.

Applying  Proposition 1 for the considered values of $S=S_m=S_*$,
$\eps_s$ and $L_s$ from \rf{5.16}--\rf{5.18} respectively we get the
"magnetic" analog of the "electric" solutions \rf{4.1}--\rf{4.a2}
\bear{5.19}
g=U_m \{g^0+\sum_{i=1}^n U_{i,m} g^i\}, \\
\label{5.20}
U_m\equiv\left(\prod_{a\in\tri_m}\prod_{I\in\Omega_{a,m}}
H_{a,m,I}^{-2\eps_g\theta_a\eps(I)d(I)
\nu_{a,m,I}^2}\right)^{\frac1{2-D}}, \\
\label{5.21}
U_{i,m}\equiv \prod_{a\in\tri_m} \prod_{I\in\Omega_{a,m}, \ I \ni i}
H_{a,m,I}^{-2\eps_g\theta_a\eps(I)\nu_{a,m,I}^2}, \\
\label{5.22}
\varphi^\beta =\varphi_m^\beta
\equiv  - \eps_g \sum_{a\in\tri_m}
\theta_a\sum_{I\in\Omega_{a,m}} \lambda_a^\beta \eps(I)
\nu_{a,m,I}^2\ln H_{a,m,I}, \\
\label{5.23}
F^a=F^{a,m}=\sum_{I\in\Omega_{a,m}} \nu_{a,m,I}
(*_0 dH_{a,m,I}) \wedge \tau(\bar I), \quad a\in\tri_m, \\
\label{5.24}
F^a = 0, \quad a\in\tri\setminus\tri_m,
\ear
where $i=1,\dots,n$; $H_{a,m,I}=H_{a,m,I}(x)>0$ are harmonic functions on
$(M_0,g^0)$
\beq{5.25}
\tri[g^0]H_{a,m,I}=0,
\eeq
and parameters satisfy the relations
\bear{5.26}
\eps_g\eps(I)\theta_a\nu_{a,m,I}^{-2}=
d(I)+\frac{(d(I))^2}{2-D}+
C^{\alpha\beta} \lambda_{\alpha a} \lambda_{\beta b}, \\
\label{5.27}
d(I\cap J)+\frac{d(I)d(J)}{2-D}+C^{\alpha\beta}\lambda_{\alpha a}
\lambda_{\beta b}=0, \quad (a,I)\ne(b,J),
\ear
$a,b\in\tri_m$; $I\in\Omega_{a,m}$; $J\in\Omega_{b,m}$.

In \rf{5.23}
\beq{5.28}
\bar I \equiv I_0 \setminus I, \quad I_0 \equiv \{1,\dots,n\}
\eeq
is "dual" set, and
$*_0d\Phi$ is the Hodge dual form on $(M_0,g^0)$ $(*_0=*[g^0])$.

The relation \rf{5.23} follows from the formula  (see (\ref{5.7}))
\beq{5.30}
F^{a,m,I}=\eps(I)\mu(I)
\exp(f- 2\sum_{i\in I}d_i\phi^i - 2 \lambda_a(\varphi))
(*_0 d\Phi^{a,m,I})\wedge\tau(\bar I),
\eeq
$f=0$ and the relation
\beq{5.31}
\exp\left(-2\lambda_a(\varphi)-2\sum_{i\in I}d_i\phi^i\right)=
H_{a,m,I}^2.
\eeq
In \rf{5.30} $\mu(I)=\pm1$ is defined by relation
\beq{5.32}
\mu(I) dx^{\mu} \wedge \tau(I_0) =
\tau(\bar I) \wedge dx^{\mu} \wedge \tau(I).
\eeq
Here we define $\nu_{a,m,I}=-\eps(I)\mu(I)\nu_s$, $s=(a,m,I)$. The
relation \rf{5.31} is a special case of a more general identity for the
solutions satisfying Proposition 1:  \beq{5.33}
\exp(2L_{As}\sigma^A)=H_s^2, \quad s\in S_*.
\eeq

Note that, for positive definite matrix $(C_{\alpha\beta})$ and $d_0\ge2$,
equation \rf{5.26} implies
\beq{5.34}
\eps(I)= \theta_a \eps_g
\eeq
for $I\in\Omega_{a,m}$; $a\in\tri_m$.
So,  for $\theta_a = 1$
Euclidean magnetically charged $p$-branes "may
live" in space-times with even number of time directions.

\subsection{Energy-momentum tensor and constraints}

We rewrite the relation \rf{5.7}
\beq{5.35}
F^{a,m,I}= \e^{-2\lambda_a(\varphi)} * \hat{F}^{a,m,I},
\quad \hat{F}^{a,m,I} = d\Phi^{a,m,I}\wedge\tau(I),
\eeq
$a \in \tri_m$. Analogously to \rf{4.11} we get
\beq{5.37}
\hat{F}^{a,m,I} \hat{F}^{a,m,J}=0 \Longrightarrow
F^{a,m,I}F^{a,m,J}=0
\eeq
for $I\ne J; I, J \in \Omega_{a,m}$ (see \rf{5.14}).

For the "composite" field $F^{a,m} = dA^{a,m}$,
$a\in\tri_m$, satisfying \rf{5.4} and \rf{5.7} we
have  (see (\ref{5.37}))
\beq{5.38}
(F^{a,m})^2=\Bigl(\sum_{I\in\Omega_{a,m}}F^{a,m,I}\Bigr)^2=
\sum_{I\in\Omega_{a,m}}(F^{a,m,I})^2
\eeq
and
\beq{5.39}
(F^{a,m}\cdot F^{a,m})_{MN}=
\sum_{I\in\Omega_{a,m}}(F^{a,m,I}\cdot F^{a,m,I})_{MN}+
\sum_{I,J\in\Omega_{a,m}\atop I\ne J}(F^{a,m,I}\cdot F^{a,m,J})_{MN}.
\eeq

The relations \rf{5.38} and \rf{5.39} imply the following relations for
the energy-momentum tensor corresponding to $F^{a,m}$ (see \rf{2.9})
\beq{5.40}
T_{MN}[F^{a,m},g]=\sum_{I\in\Omega_{a,m}}T_{MN}[F^{a,m,I},g]+
\bar T_{MN}[F^{a,m},g],
\eeq
where
\beq{5.41}
\bar T_{MN}[F^{a,m},g]\equiv\frac1{(n_a-1)!}
\sum_{I,J\in\Omega_{a,m}\atop I\ne J}(F^{a,m,I}\cdot F^{a,m,J})_{MN}.
\eeq

Analogously to \rf{4.18}, \rf{4.19} we get
\beq{5.42}
T_{MN}[\hat{F}^{a,m},g]=\sum_{I\in\Omega_{a,m}}T_{MN}[\hat{F}^{a,m,I},g]+
\bar T_{MN}[\hat{F}^{a,m},g],
\eeq
where
\beq{5.43}
\bar T_{MN}[\hat{F}^{a,m},g]\equiv\frac1{(n_a^{*} -1)!}
\sum_{I,J\in\Omega_{a,m}\atop I\ne J}(\hat{F}^{a,m,I}\cdot
\hat{F}^{a,m,J})_{MN},
\eeq
where $n_a^{*} = D - n_a$ and  $a\in\tri_m$. Here
\bear{5.44}
\hat{F}^{a,m}= 0, \quad a\in\tri\setminus\tri_m, \\
\label{5.45}
\hat{F}^{a,m}=\sum_{I\in\Omega_{a,m}}\hat{F}^{a,m,I}, \quad a\in\tri_m.
\ear

It is clear that
\beq{5.46}
F^{a,m}=
\e^{-2\lambda_a(\varphi)} * \hat{F}^{a,m}.
\eeq

>From \rf{5.13}, \rf{5.40}-\rf{5.46} we get
\bear{5.47}
T_{MN}[F^{a,m},g]
= - \eps_g \e^{-4\lambda_a(\varphi)}  T_{MN}[\hat{F}^{a,m},g], \\
\label{5.48}
\bar T_{MN}[F^{a,m},g]
= - \eps_g \e^{-4\lambda_a(\varphi)} \bar T_{MN}[\hat{F}^{a,m},g].
\ear

Using the results from the previous section
applied to $T_{MN}[\hat{F}^{a,m},g]$
and \rf{5.47}, \rf{5.48}
we obtain that the non-zero
components for $\bar T_{MN}$ may take place only if the condition \rf{4.13}
holds and in this case
\bear{5.49}
T_{1_i1_j}[F^{a,m},g]=\bar T_{1_i1_j}[F^{a,m},g] \nn \\
=\sqrt{|g^i|}\sqrt{|g^j|}\exp(-2\gamma -4\lambda_a(\varphi)) (- \eps_g)
C_{ij}(\Phi^{a,m},\phi,g^0),
\ear
where $C_{ij}(\Phi^{a,m},\phi,g^0)$ is defined in (\ref{4.21}) (with $o =
m$).

The non-block-diagonal part of the total energy-momentum tensor \rf{2.7}
(in magnetic case) has the following form
\beq{5.51}
T_{1_i1_j}[F^m,\varphi,g]=\sqrt{|g^i|}\sqrt{|g^j|}
\e^{-2\gamma}C_{ij}^m(\Phi^m,\varphi,\phi,g^0),
\eeq
where  $\Phi^{m} = (\Phi^{a,m})$ and
\beq{5.52}
C_{ij}^{m}(\Phi^{m},\varphi,\phi,g^0)\equiv (- \eps_g)
\sum_{a\in\tri_m}\theta_a\exp[- 2\lambda_a(\varphi)]
C_{ij}(\Phi^{a,m},\phi,g^0),
\eeq
$i,j\in w_1$; $i\ne j$. Relations \rf{5.51}, \rf{5.52} follows from
relations \rf{2.7}, \rf{5.49}.

>From Einstein-Hilbert equations \rf{2.4} $(\Lambda=0)$,
block-diagonal form of metric and Ricci-tensor (in the considered ansatz)
and \rf{5.51}
we get the magnetic analog of constraints \rf{4.28}
\beq{5.54}
C_{ij}^m(\Phi^m,\varphi,\phi,g^0)=0, \quad i<j,
\eeq
$i,j\in w_1$.

{\bf Restriction 2m.}
The constraints are satisfied identically in the case
$n_1\le1$ \rf{2.r} or when $n_1>1$ and
\beq{5.55}
W_{ij}(\Omega_{a,m})= \emptyset,
\eeq
$i<j$; $i,j\in w_1$; $a\in\tri_m$.

\section{Electro-magnetic case}
Now we consider the "superposition" of the
ans${\ddot a}$tze from Sections 4 and 5, i.e. we put
\bear{6.1}
F^a=0,\quad a\in\tri\setminus(\tri_e\cup\tri_m), \\
\label{6.2}
F^a=F^{a,e}=\sum_{I\in\Omega_{a,e}}F^{a,e,I},\quad
a\in\tri_e\setminus\tri_m, \\
\label{6.3}
F^a=F^{a,m}=\sum_{I\in\Omega_{a,m}}F^{a,m,I},\quad
a\in\tri_m\setminus\tri_e, \\
\label{6.4}
F^a=F^{a,e}+F^{a,m}=\sum_{I\in\Omega_{a,e}}F^{a,e,I}+
\sum_{J\in\Omega_{a,m}}F^{a,m,J},\quad a\in\tri_e\cap\tri_m,
\ear
where  $F^{a,e,I}$ and  $F^{a,m,J}$ are defined in
(\ref{2.21}) and (\ref{5.7}) respectively.

\subsection{Energy-momentum tensor and constraints}

Let $d_0\ne2$. For $a\in\tri_e\cap\tri_m$ we obtain
\beq{6.5}
(*\hat F^{a,m,I})F^{a,e,J}=F^{a,e,J}(*\hat F^{a,m,I})=0,
\eeq
$I\in\Omega_{a,m}$; $J\in\Omega_{a,e}$; and hence
\beq{6.6}
F^{a,e}F^{a,m}=F^{a,m}F^{a,e}=0.
\eeq
Relation \rf{6.5} is due to the non-equal numbers of $M_0$
indices in the non-zero  components of forms
$\hat F$ and $F$ in \rf{6.5}. From \rf{6.6} we get
\beq{6.7}
(F^a)^2=(F^{a,e})^2+(F^{a,m})^2,
\eeq
where $(F^{a,e})^2$ and $(F^{a,m})^2$ are expressed by \rf{4.16} and
\rf{5.38}. We also get
\bear{6.8}
(F^a\cdot F^a)_{MN}=(F^{a,e}\cdot F^{a,e})_{MN}+
(F^{a,m}\cdot F^{a,m})_{MN} \nn \\
+(F^{a,e}\cdot F^{a,m})_{MN}+(F^{a,m}\cdot F^{a,e})_{MN}
\ear
for $a\in\tri_e\cap\tri_m$.

The relations \rf{4.16}, \rf{5.38}, \rf{6.7}, \rf{6.8} imply
\beq{6.9}
T_{MN}[F^a,g]=T_{MN}[F^{a,e},g]+T_{MN}[F^{a,m},g]+
\til T_{MN}[F^{a,e},F^{a,m},g],
\eeq
where $T_{MN}[F^{a,e},g]$, $T_{MN}[F^{a,m},g]$ are presented in \rf{4.18},
\rf{5.40} respectively and
\beq{6.10}
\til T_{MN}^a=\til T_{MN}[F^{a,e},F^{a,m},g]\equiv
\frac1{(n_a-1)!}\bigl\{(F^{a,e}\cdot F^{a,m})_{MN}+
(F^{a,m}\cdot F^{a,e})_{MN}\bigr\},
\eeq
$a\in\tri_e\cap\tri_m$. We obtain
\bear{6.11}
(F^{a,m} \cdot F^{a,e})_{MN}=(F^{a,e}\cdot F^{a,m})_{NM} \nn \\
=\sum_{I\in\Omega_{a,m}}\sum_{J\in\Omega_{a,e}}
\e^{-2\lambda_a(\varphi)}[(*\hat F^{a,m,I})\cdot F^{a,e,J}]_{MN},
\ear
$a\in\tri_e\cap\tri_m$.

The tensor \rf{6.10} is trivial for $d_0>3$ and may have non-zero
(non-diagonal) components for $d_0=1,3$, when $n_1=|w_1|\ge1$ (i.e. there
are $1$-dimensional $M_i$).

Indeed, calculations give (for $d_0\ne2$) that
\beq{6.12}
[(*\hat F^{a,m,I})\cdot F^{a,e,J}]_{MN}=
[F^{a,e,J}\cdot(*\hat F^{a,m,I})]_{NM}
\eeq
may have non-zero components only if $d_0=1$; $\bar I=\{j\}\sqcup J$,
$j\in w_1$ :
\bear{6.13}
[(*\hat F^{a,m,I})\cdot F^{a,e,J}]_{1_j\mu}=
[F^{a,e,J}\cdot(*\hat F^{a,m,I})]_{\mu1_j}=
d(J)!\eps_g\mu(I)\delta(j,J)|g^j|^{1/2} \nn \\
\times\exp\Bigl(-3\gamma-\sum_{i=1}^nd_i\phi^i+2\phi^j\Bigr)
|g^0|^{-1/2}\p_{1_0}\Phi^{a,m,I}\p_\mu\Phi^{a,e,J},
\ear
or $d_0=3$; $J=\{i\}\sqcup\bar I$, $i\in w_1$
\bear{6.14}
[(*\hat F^{a,m,I})\cdot F^{a,e,J}]_{\mu1_i}=
[F^{a,e,J}\cdot(*\hat F^{a,m,I})]_{1_i\mu}=
\eps_{g_0}\eps_g\delta(i,I)\mu(I)d(J)! \nn \\
\times|g^i|^{1/2}\bigl(|g^0|^{1/2}
\eps_{\mu\rho\nu}\nabla^\rho\Phi^{a,m,I}\nabla^\nu\Phi^{a,e,J}\bigr)
\exp\Bigl(-\gamma-\sum_{i=1}^nd_i\phi^i\Bigr).
\ear
Recall that $w_1$ is defined in \rf{4.12}.

>From \rf{6.9}--\rf{6.14} we have for $d_0=1$
\beq{6.15}
T_{1_j1_0}=\eps_g|g^j|^{1/2}|g^0|^{-1/2}
\exp\Bigl(-3\gamma-\sum_{i=1}^nd_i\phi^i+2\phi^j\Bigr)
C_{j1_0}^{(1)}(\Phi^e,\Phi^m),
\eeq
where
\beq{6.16}
C_{j1_0}^{(1)}(\Phi^e,\Phi^m)\equiv\sum_{a\in\tri_e\cap\tri_m}
\sum_{(I,J) \in W_j^{(1)}}
\mu(I)\delta(j,J)\p_{1_0}\Phi^{a,m,I}\p_{1_0}\Phi^{a,e,J},
\eeq
and
\beq{6.17}
W_j^{(1)}= W_j^{(1)}(\Omega_{a,m},\Omega_{a,e}) \equiv
\{(I,J)\in\Omega_{a,m}\times\Omega_{a,e}|\bar I=\{j\}\sqcup J\},
\eeq
$j\in w_1$. For $d_0=3$ the analogous relations read
\bear{6.18}
T_{\mu1_i}=\eps_{g_0}\eps_g|g^i|^{1/2}
\exp\Bigl(-\gamma-\sum_{i=1}^nd_i\phi^i\Bigr)
C_{i\mu}^{(3)}(\Phi^e,\Phi^m,g^0); \\
\label{6.19}
C_{i\mu}^{(3)}(\Phi^e,\Phi^m,g^0)\equiv\sum_{a\in\tri_e\cap\tri_m}
\sum_{(I,J)\in W_i^{(3)}}
\delta(i,I)\mu(I)|g^0|^{1/2} \nn \\
\times\eps_{\mu\rho\nu}\nabla^\rho\Phi^{a,m,I}
\nabla^\nu\Phi^{a,e,J}; \\
\label{6.20}
W_i^{(3)} = W_i^{(3)}(\Omega_{a,m},\Omega_{a,e}) \equiv
\{(I,J)\in\Omega_{a,m}\times\Omega_{a,e}|J=\{i\}\sqcup\bar I\},
\ear
$i\in w_1$ and $\mu = 1_0,2_0,3_0$.

Thus, in the "electro-magnetic" case for $d_0=1,3$ we are led to
$n_1=|w_1|$ additional constraints
\beq{6.21}
C_{i\mu}^{(d_0)}(\Phi^e,\Phi^m,g^0)=0,
\eeq
$i\in w_1$  and $\mu = 1_0, \ldots ,(d_0)_0$.

As to $T_{1_i1_j}$ -- components for $i,j\in w_1$ we get from \rf{6.9},
\rf{4.20}, \rf{4.21}, \rf{5.49}
\beq{6.22}
T_{1_i1_j}=\sqrt{|g^i|}\sqrt{|g^j|}\e^{-2\gamma}
C_{ij}(\Phi^e,\Phi^m,\varphi,\phi,g^0),
\eeq
where
\bear{6.23}
C_{ij}(\Phi^e,\Phi^m,\varphi,\phi,g^0)=\sum_{a\in\tri_e}\theta_a
\exp[2\lambda_a(\varphi)]C_{ij}(\Phi^{a,e},\phi,g^0) \nn \\
-\eps_g\sum_{a\in\tri_m}\theta_a
\exp[-2\lambda_a(\varphi)]C_{ij}(\Phi^{a,m},\phi,g^0)
\ear
(see \rf{4.21}).

Thus, for $n_1\ge2$ the constraints \rf{4.28}, \rf{5.54} are generalized as
follows
\beq{6.24}
C_{ij}(\Phi^e,\Phi^m,\varphi,\phi,g^0)=0,\quad i<j,
\eeq
$i,j\in\ w_1$.

Now let us consider the case $d_0=2$. It may be verified that all the
presented in this section relations are unchanged if the following
restrictions are imposed:
\bear{6.25}
W(\Omega_{a,e},\hat\Omega_{a,m})=\emptyset, \\
\label{6.26}
W_{ij}(\Omega_{a,e},\hat\Omega_{a,m})=\emptyset,
\ear
where $i \neq j$; $i,j\in w_1$, $a\in\tri_e\cap\tri_m$ and
\bear{6.27}
\hat\Omega_{a,m}=\{I\in\Omega|\bar I\in\Omega_{a,m}\}, \\
\label{6.28}
W(\Omega_1,\Omega_2)\equiv
\{(I_1,I_2)\in\Omega_1\times\Omega_2|I_1=I_2\}, \\
\label{6.28a}
W_{ij}(\Omega_{1}, \Omega_{2}) \equiv
\{(I_1,I_2)\in\Omega_1\times\Omega_2
|I_1 =\{i\}\sqcup(I_1 \cap I_2), \
I_2=\{j\}\sqcup(I_1 \cap I_2) \}.
\ear

\subsection{$\sigma$-model}

In the general case we are led to the $\sigma$-model \rf{2.29}-\rf{2.30b}
with
\beq{6.29}
{\cal L}_A={\cal L}_{A,e}+{\cal L}_{A,m},
\eeq
where ${\cal L}_{A,e}$ and ${\cal L}_{A,m}$ are defined in \rf{2.30} and
\rf{5.11} respectively. We also obtain constraints \rf{6.24} for
$n_1\ge2$, $d_0\ne2$ and \rf{6.21} for $n_1\ge1$ for $d_0=1,3$. For $d_0=2$
the constraints are the same if the restrictions \rf{6.25}, \rf{6.26}
hold.

We recall that all the constraints occur due to non-block-diagonal part of
energy-momentum tensor. The block-diagonal part gives rise to
$\sigma$-model itself.

Thus we are led to the following

{\bf Proposition 2.} Let us consider
the model (\ref{2.1}) where the manifold, metric,
scalar fields  and forms  are defined by relations
(\ref{2.10}), (\ref{2.11})-(\ref{2.13}), (\ref{2.24}) and
(\ref{6.1})-(\ref{6.4}) respectively.
Then  for $d_0 \ne 2$  and
$\gamma  = {\gamma}_{0}(\phi)$   from (\ref{3.1})
the equations of motion (\ref{2.4})-(\ref{2.6})
and Bianchi identities (\ref{5.9})  are equivalent
to the equations of motion for the $\sigma$-model
(\ref{3.2})-(\ref{3.4}) with the Lagrangians
${\cal L}$, ${\cal L}_{\varphi}$ from (\ref{2.30a}), (\ref{2.30b}) and
${\cal L}_A$ from (\ref{6.29})  (see also (\ref{2.30}) and (\ref{5.11}))
and  the constraints (\ref{6.24}) (for all $d_0 \ne 2$ )
and (\ref{6.21}) (for $d_0 = 1,3$) imposed.

{\bf Proof.} The appearance of constraints was verified above.
Now we consider the reduction to $\sigma$-model itself.
For  $(F,\varphi)$-part of field equations and Bianchi identities
the equivalence with corresponding equations of motion for
$\sigma$-model can be readily verified.
Here  we consider  the Einstein equations (\ref{2.4}) written
in the form
\beq{6.30n}
R_{MN}  =   Z_{MN} + \frac{2\Lambda}{D - 2} g_{MN},
\eeq
where
\beq{6.31n}
Z_{MN} \equiv   T_{MN} + \frac{T}{2 -D} g_{MN},
\eeq
and $T = {T_M}^M$. Here
\beq{6.32n}
Z_{MN} =   Z_{MN}[\varphi]
+ \sum_{a\in\Delta} \theta_a  e^{2 \lambda_{a}(\varphi)} Z_{MN}[F^a,g],
\eeq
where
\bear{6.33n}
Z_{MN}[\varphi] =
C_{\alpha\beta} \p_{M} \varphi^{\alpha} \p_{N} \varphi^{\beta},
\\
Z_{MN}[F^a,g] = \frac{1}{n_{a}!}  \left[ \frac{n_a -1}{2 -D}
g_{MN} (F^{a})^{2}
 + n_{a}  F^{a}_{M M_2 \ldots M_{n_a}} F_{N}^{a, M_2 \ldots M_{n_a}}
 \right].
\label{6.34n}
\ear
For block-diagonal part of  (\ref{6.34n}) we have
>from (\ref{6.7}) and (\ref{6.8})
(see also (\ref{4.16}),(\ref{4.17}),(\ref{5.38})
and (\ref{5.39}))
\bear{6.35n}
Z_{MN}[F^a,g] =
\sum_{I \in \Omega_{a,e}} Z_{MN}[F^{a,e,I},g]+
\sum_{J\in\Omega_{a,m}} Z_{MN}[F^{a,m,J},g],
\ear
where  $(M,N)= (\mu,\nu), (m_i,n_i)$; $i=1, \ldots, n$
($F^{a,e,I}$ and  $F^{a,m,J}$ are defined in
(\ref{2.21}) and (\ref{5.7}) respectively).
Here we put $\Omega_{a,e} = \emptyset$ and
$\Omega_{b,m} = \emptyset$ for $a \notin \tri_e$  and
$b \notin \tri_m$ respectively.
Using the relations for Ricci tensor
(\ref{2.25}), (\ref{2.26})  with
$\gamma = {\gamma}_{0}(\phi)$   from (\ref{3.1})
and relations (\ref{2.38}), (\ref{4.4}), (\ref{4.5}),
(\ref{5.14}),(\ref{5.15}) and (\ref{6.4}) we
obtain that $(m_i,n_i)$-components of Einstein equations (\ref{6.30n})
($i=1, \ldots, n$)
are equivalent to $\phi^i$-part of $\sigma$-model equations
and $(\mu,\nu)$-components of Einstein equations (\ref{6.30n})
are equivalent to $g^0$-part of $\sigma$-model equations of motion
(or $\sigma$-model Einstein equations). Note
that dealing with  $(\mu,\nu)$-components of (\ref{6.30n})
we use the relation for $\gamma = {\gamma}_{0}(\phi)$  with  $\phi$
substituted from $(m_i,n_i)$-equations. Also the following  relations
should be used:
\beq{5.13n}
Z_{MN}[*F,g]=-\eps_g Z_{MN}[F,g]
\eeq
(see \rf{5.13}) and
\beq{6.36n}
G^{ij} d_{j} =  \frac{2 - d_0}{2-D}, \quad
G^{ij} l_{j I}=  - \sum_{k \in I} \delta^i_k
+ \frac{d(I)}{D-2},
\eeq
where  $l_{j I} $ are defined in (\ref{3.8}). The proposition
is proved.

{\bf Remark 2.} We may also fix the gauge
$\gamma = {\gamma}(\phi)$  (${\gamma}(\phi)$ is smooth function)
by arbitrary manner or do not fix it. In this case the
Proposition 2 is simply modified by the replacement of the action
(\ref{3.2}) by the action (\ref{2.29})-(\ref{2.30b}) with
${\cal L}_A$ from (\ref{6.29}). This is valid also for
$d_0 = 2$ if the restrictions (\ref{6.28}) and
(\ref{6.28a}) are imposed.

\subsection{Exact solutions}
When $\xi_i=\Lambda=0$ and all the  above constraints are
satisfied we deal with $\sigma$-model \rf{3.5}, where
\beq{6.30}
S=S_e\sqcup S_m,
\eeq
$\eps_s$ are defined in \rf{3.e}, \rf{5.17} and $L_s=(L_{As})$ are defined
in \rf{3.7} and \rf{5.18} for $s\in S_e$ and $s\in S_m$ respectively.

Using the Proposition 1 we obtain the exact solutions generalizing pure
electric and magnetic ones:
\bear{6.31}
g=U\left\{g^0+\sum_{i=1}^{n} U_i g^i\right\}, \\
\label{6.32}
U=U_eU_m,\quad U_i= U_{i,e}U_{i,m}, \\
\label{6.33}
\varphi^\beta=\varphi_e^\beta+\varphi_m^\beta,
\ear
where $U_e$, $U_{i,e}$, $U_m$, $U_{i,m}$, $\varphi_e$, $\varphi_m$ are
presented in \rf{4.1e}, \rf{5.1i}, \rf{5.20}, \rf{5.21}, \rf{4.p},
\rf{5.22} respectively. The fields of forms are given by
\rf{6.1}--\rf{6.4}, where
\bear{6.34}
F^{a,e,I}=\nu_{a,e,I}dH_{a,e,I}^{-1}\wedge\tau(I), \\
\label{6.35}
F^{b,m,J}=\nu_{b,m,J} (*_0 dH_{b,m,J}) \wedge\tau(\bar J),
\ear
$I\in\Omega_{a,e}$; $J\in\Omega_{b,m}$; $a\in\tri_e$; $b\in\tri_m$
($\bar J$ is defined in \rf{5.28})
and $*_0 dH$ is the Hodge dual form on $(M_0,g^0)$.

Here
parameters $\nu_{a,e,I}$ and $\nu_{b,m,J}$ satisfy the relations \rf{4.2},
\rf{5.26} respectively. The dimensions (of "branes") $d(I)$ and $\lambda_a$
satisfy relations \rf{4.3}, \rf{5.27} and the following crossing
orthogonality relation
\beq{6.36}
d(I\cap J)+\frac{d(I)d(J)}{2-D}-\lambda_a\cdot\lambda_b=0,
\eeq
$I\in \Omega_{a,e}$; $J\in \Omega_{b,m}$; $a\in\tri_e$; $b\in\tri_m$
corresponding to \rf{3.12} with $s =(a,e,I)\in S_e$,
$r = (b,m,J) \in S_m$.

All functions $H_{a,e,I}$, $H_{b,m,J}$ are harmonic on $(M_0,g^0)$:
\beq{6.37}
\tri[g^0]H_{a,e,I}=0,\quad \tri[g^0]H_{b,m,J}=0,
\eeq
$I\in\Omega_{a,e}$; $J\in\Omega_{b,m}$; $a\in\tri_e$; $b\in\tri_m$.

In more compact form relations \rf{4.2}, \rf{4.3}, \rf{5.26}, \rf{5.27},
\rf{6.36} read
\beq{6.38}
d(I_s \cap I_r)+ \frac{d(I_s) d(I_r)}{2-D}+
\chi_s \chi_r \lambda_{a_s} \cdot\lambda_{a_r}=
-\eps_s \nu_s^{-2}\delta_{sr},
\eeq
$s,r \in S$. Here we denote $s=(a_s,o_s,I_s)$,
$\nu_s=\nu_{a_s,o_s,I_s}$; $o_s= e,m$;
$\chi_s=+1,-1$ for  $s \in S_e, S_m$ respectively;
$I_s \in \Omega_{a_s,o_s}$, $a_s \in \tri_{o_s}$, $s \in S$.

The solutions  (\ref{6.31})-(\ref{6.38})
are valid if the restrictions   (\ref{4.30}), (\ref{5.55}),
\beq{6.38a}
W_i^{(d_0)}(\Omega_{a,m},\Omega_{a,e}) = \emptyset,
\eeq
$i\in w_1$ for $d_0 = 1,3$ (see (\ref{6.17}), (\ref{6.20}))
and (\ref{6.28}), (\ref{6.28a}) for $d_0 = 2$ are imposed.

\subsubsection*{\bf $D= 12$ model }

Here we illustrate the obtained above general solution
by considering a bosonic field model in dimension  $D= 12$ \cite{KKP}
that admits the bosonic sector of 11-dimensional
supergravity as a consistent truncation.
The action for this model  with omitted
Chern-Simons term has the following form
\beq{a.1}
\hat{S}_{12} =
\int_{M} d^{12}z \sqrt{|g|} \{ {R}[g] -
g^{MN} \partial_{M} \varphi \partial_{N} \varphi
- \frac{1}{4!} \exp( 2 \lambda_{1} \varphi)  (F^1)^2
- \frac{1}{5!} \exp( 2 \lambda_{2} \varphi ) (F^2)^2 \}.
\eeq
Here  ${\rm rank} F^1 = 4$, ${\rm rank F^2} = 5$,  and
\beq{a.2}
 \lambda_{1}^2 = - \frac{1}{10}, \quad
 \lambda_{2} = - 2 \lambda_{1}.
\eeq
In \rf{2.1} $\Delta$ = \{1, 2 \} and  all $\theta_a = 1$, $a=1,2$.
We put
$\Delta_e  =  \Delta_m = \Delta$, i.e. for both
two forms we consider the composite ansatz with
electric and magnetic components (see \rf{6.4}).

The dimensions of p-brane worldsheets are
\bear{a.3}
d(I) = &&3, \qquad  I \in  \Omega_{1,e},   \\ \nn
       &&7, \qquad  I \in  \Omega_{1,m},  \\ \nn
       &&4, \qquad  I \in  \Omega_{2,e},  \\ \nn
       &&6, \qquad  I \in  \Omega_{2,m}
\ear
(see \rf{2.23} and \rf{5.2}).  Thus, the model describes
electrically charged 2- and 3-branes and magnetically
charged 6- and 5-branes (corresponding to $F^1$ and $F^2$
respectively).

>From relations \rf{6.38} we obtain the intersection rules:
\beq{a.4}
\begin{array}{rll}\displaystyle
d(I \cap J) = &1, \ \{ d(I), d(J)\} =& \{ 3, 3\}, \{ 3, 4\},  \\
              &2,                   & \{ 3, 6\},
                  \{ 3, 7\}, \{ 4, 4\},\{ 4, 6\},  \\
              &3,                   & \{ 4, 7\},   \\
              &4,                   & \{ 6, 6\}, \{ 6, 7\},   \\
              &5,                   & \{ 7, 7\},
              \end{array}
\eeq
and
\beq{a.5}
 \nu^{2}_{1,e,I} = \nu^{2}_{1,m,I}=
 \nu^{2}_{2,e,I}= \nu^{2}_{2,m,I} = \frac{1}{2}
\eeq
for all $I$. Also we get
\bear{a.6}
\eps(I) = -1, \quad I \in  \Omega_{1,e} \cup \Omega_{2,e},   \\
\eps(J) = \eps_g, \quad J \in  \Omega_{1,m} \cup \Omega_{2,m},
\label{a.7}
\ear
(recall that $\eps_g \equiv \sign\det(g_{MN})$).
Thus, electrically charged $p$-branes should have
odd number of time directions and
magnetically charged $p$-branes should have
even number of time directions for $\eps_g = 1$
and odd number of time directions for $\eps_g = -1$.
Relations \rf{a.4} are intersection rules for $p$-branes and
relations \rf{a.6}, \rf{a.7} are signature restrictions on them.
We note that due to relations \rf{a.4} all constraints treated
in previous sections are satisfied identically.

The metric \rf{6.31} reads
\bear{a.8}
g= U_e U_m \biggr\{g^0 + \sum_{i=1}^{n}
U_{i,e}U_{i,m} g^i \biggr\}, \\
\label{a.9}
U_e =   \biggr( \prod_{I_1 \in \Omega_{1,e}}
H_{1,e,I_1} \biggr)^{\frac{3}{10}}
\biggr( \prod_{I_2 \in \Omega_{2,e}}
H_{2,e,I_2} \biggr)^{\frac{2}{5}},   \\
\label{a.10}
U_m =   \biggr( \prod_{J_1 \in \Omega_{1,e}}
H_{1,m,J_1} \biggr)^{\frac{7}{10}}
\biggr( \prod_{J_2 \in \Omega_{2,m}}
H_{2,m,J_2} \biggr)^{\frac{3}{5}},  \\
\label{a.11}
U_{i,e} =
\biggr( \prod_{I_1 \in \Omega_{1,e}, \ I_1 \ni i }
H_{1,e,I_1}^{-1} \biggr)
 \prod_{I_2 \in \Omega_{2,e}, \ I_2 \ni i }
H_{2,e,I_2}^{-1},  \\
\label{a.12}
U_{i,m} =
\biggr( \prod_{J_1 \in \Omega_{1,m}, \ J_1 \ni i }
H_{1,m,J_1}^{-1} \biggr)
 \prod_{J_2 \in \Omega_{2,m}, \ J_2 \ni i }
H_{2,m,J_2}^{-1}.
\ear
The scalar field is
\bear{a.13}
\varphi =
&\sum_{I_1 \in \Omega_{1,e}} \eta \ln H_{1,e,I_1}
-2 \sum_{I_2 \in \Omega_{2,e}} \eta \ln H_{2,e,I_2}  \\ \nn
&- \sum_{J_1 \in \Omega_{1,m}} \eta \ln H_{1,m,J_1}
+ 2 \sum_{J_2 \in \Omega_{2,m}} \eta \ln H_{2,m,J_2},
\ear
where $\eta = \lambda_1/2$.
The fields of forms are the following
\bear{a.14}
F^{1}= \sum_{I_1 \in \Omega_{1,e}} \nu_{1,e,I_1}
 dH_{1,e,I_1}^{-1} \wedge \tau(I_1) +
\sum_{J_1 \in \Omega_{1,m}} \nu_{1,m, J_1}
(*_0 dH_{1,m,J_1}) \wedge \tau(\bar{J}_1), \\
\label{a.15}
F^{2}= \sum_{I_2 \in \Omega_{2,e}} \nu_{2,e,I_2}
 dH_{2,e,I_2}^{-1} \wedge \tau(I_2) +
\sum_{J_2 \in \Omega_{2,m}} \nu_{2,m,J_2}
(*_0  dH_{2,m,J_2}) \wedge \tau(\bar{J}_2),
\ear
where
$*_0  dH$ is the Hodge dual form on $(M_0,g^0)$. The metric and all fields
are defined on manifold \rf{2.10} and all functions
$H_{a,o,I}$  are  harmonic on $M_0$.

The subsets  $\Omega_{a,o} \in \Omega$ satisfy the
intersection rules \rf{a.4}  and
the  signature restrictions \rf{a.6}, \rf{a.7}.

These solutions also satisfy the equations
of motion  for $D = 12$ model from ref. \cite{KKP}
with Chern-Simons term included. This can be readily
verified using the relation for the total action
\beq{a.16}
S_{12} =  \hat{S}_{12} + c_{12} \int_{M} A^2 \wedge F^1  \wedge F^1
\eeq
where $c_{12} = {\rm const}$ and $\hat{S}_{12}$ is defined in \rf{a.1}
($F^2  =d A^2$).
Note that the elementary p-brane solutions corresponding to  $F^1$ field
($p =2,6$) were considered in \cite{KKP}.

\subsubsection*{\bf $D= 11$ supergravity }

Now we consider as another example the $D= 11$ supergravity \cite{CJS,SS}.
The action for the bosonic sector of this
theory  with omitted  Chern-Simons term has the
following form
\beq{b.1}
\hat{S}_{11} =
\int_{M} d^{11}z \sqrt{|g|} \{ {R}[g] -  \frac{1}{4!}  (F)^2 \}.
\eeq
Here  ${\rm rank} F = 4$.

The dimensions of p-brane worldsheets are
(see (\ref{2.23}) and (\ref{5.2}))
\bear{b.2}
d(I) = &&3, \quad  I \in  \Omega_{e},   \\ \nn
       &&6, \quad  I \in  \Omega_{m}.
\ear
The model describes
electrically charged 2-branes and magnetically charged  5-branes.

>From \rf{6.38} we obtain the intersection rules \cite{Ts1}
\beq{b.4}
\begin{array}{rll}\displaystyle
d(I \cap J) = &1, \  \{ d(I), d(J)\} =& \{ 3, 3\},   \\
              &2,                     & \{ 3, 6\},    \\
              &4,                     & \{ 6, 6\},
              \end{array}
\eeq
and
\beq{b.5}
 \nu^{2}_{e,I}= \nu^{2}_{m,I} = \frac{1}{2}
\eeq
for all $I$. Here and below we omitted the index "1" numerating
forms. Also we get
\bear{b.6}
\eps(I) = -1, \quad I \in  \Omega_{e},    \\
\eps(J) = \eps_g, \quad J \in  \Omega_{m}.
\label{b.7}
\ear
The stress-tensor restrictions are also satisfied for these solutions.

The metric \rf{6.31} reads
\bear{b.8}
g= U_e U_m \biggr\{g^0 + \sum_{i=1}^{n}
U_{i,e}U_{i,m} g^i \biggr\}, \\ \label{b.9}
U_e =
\biggr( \prod_{I \in \Omega_{e}}
H_{e,I} \biggr)^{\frac{1}{3}}, \qquad
U_m =
\biggr( \prod_{J \in \Omega_{e}}
H_{m,J} \biggr)^{\frac{2}{3}}, \\ \label{b.11}
U_{i,e} =
 \prod_{I \in \Omega_{e}, \ I \ni i }
H_{e,I}^{-1}, \qquad
U_{i,m} =
 \prod_{J \in \Omega_{m}, \ J \ni i }
H_{m,J}^{-1}.
\ear
The fields of forms are following
\beq{b.14}
F = \sum_{I \in \Omega_{e}} \nu_{e,I}
 dH_{e,I}^{-1} \wedge \tau(I) +
 \sum_{J \in \Omega_{m}} \nu_{m,J}
(*_0  dH_{m,J}) \wedge \tau(\bar{J}),
\eeq
where $*_0  dH$ is the Hodge dual form on $(M_0,g^0)$.
The metric and all fields
are defined on manifold \rf{2.10} and all functions
$H_{o,I}$, $o= e,m$, are  harmonic on $M_0$.

The subsets  $\Omega_{o} \in \Omega$, $o = e,m$, satisfy the intersection
rules \rf{b.4} and the signature restrictions \rf{b.6}, \rf{b.7}.

The solutions also satisfy  the equations
of motion  with Chern-Simon term taken into account.
This can be readily
verified using the relation for the bosonic part of action
for $D = 11$ supergravity \cite{CJS,SS}
\beq{b.16}
S_{11} =  \hat{S}_{11} + c_{11} \int_{M} A \wedge F  \wedge F
\eeq
where $c_{11} = {\rm const}$ and $\hat{S}_{11}$ is defined in \rf{b.1}
($F  =d A$). We note that these solutions (see also \cite{IM4})
coincide with those obtained in  \cite{BREJS,Ts1}) for
flat $M_{\nu}$, $\nu = 0, \ldots, n$.

\section{Generalization to non-Ricci-flat spaces}

Here we present a generalization of the  above solution to the
case of non-Ricci flat space $(M_0,g^0)$ and when some additional internal
Einstein spaces of non-zero curvature $(M_i,g^i)$, $i=n+1,\dots,n+k$, are
included.

\subsection{Non-Ricci-flat solutions for $\sigma$-model with a potential}
Let us consider the $\sigma$-model governed by the action
\bear{7.1}
S_\sigma=S_\sigma[g^0,\sigma,\Phi]=\int_{M_0}d^{d_0}x
\sqrt{|g^0|}\Bigl\{R[g^0]-
\hat G_{AB}g^{0\mu\nu} \p_\mu\sigma^A\p_\nu\sigma^B \nn
-2V(\sigma) \\
-\sum_{s\in S} \eps_s \e^{2L_{As} \sigma^A}
g^{0 \mu\nu} \p_\mu \Phi^s \p_\nu \Phi^s \Bigr \}.
\ear
Here $(\hat G_{AB})$ is non-degenerate matrix and $\eps_s\ne0$, $s \in S$
($S \neq \emptyset$).

The equations of motion for the action \rf{7.1} have the following form
\bear{7.2}
R_{\mu\nu}[g^0]=\hat G_{AB}\p_\mu\sigma^A\p_\nu\sigma^B+
\frac{2V}{d_0-2}g_{\mu\nu}^0 + \sum_{s\in S}\eps_s\e^{2L_{As}\sigma^A}
\p_\mu\Phi^s\p_\nu\Phi^s,
\\ \label{7.3}
\hat G_{AB}\tri[g^0]\sigma^B-\frac{\p V}{\p\sigma^A}-
\sum_{s\in S}\eps_s L_{As}\e^{2L_{Cs}\sigma^C}(\p\Phi^s)^2=0, \\
\label{7.4}
\p_\mu\left(\sqrt{|g^0|}g^{0\mu\nu}
\e^{2L_{As}\sigma^A}\p_\nu\Phi^s\right)=0,
\ear
$s\in S$. In what follows we consider the potential of a special form
\beq{7.5}
V=V(\sigma)=\sum_{c=1}^k A_c \exp(u_{c A}\sigma^A),
\eeq
where  $A_c \neq 0$ and
vectors $u_c=(u_{c A})$ satisfy the orthogonality
conditions
\beq{7.6}
\hat G^{AB}u_{c A}L_{Bs}=0,
\eeq
$c=1,\dots,k$; $s\in S$.

We also consider the "truncated" action
\beq{7.7}
S_{\sigma,0}=S_{\sigma,0}[g^0,\hat\sigma]=
\int_{M_0}d^{d_0}x\sqrt{|g^0|}\bigl\{R[g^0]-
\hat G_{AB}g^{0\mu\nu}\p_\mu\hat\sigma^A\p_\nu\hat\sigma^B-
2V(\hat\sigma)\bigr\},
\eeq
i.e. the action \rf{7.1} with $\Phi^s=0$, $s\in S$,
and the action \rf{7.1} with  omitted
curvature and  potential terms
\beq{7.7a}
S_{\sigma,1}=S_{\sigma,1}[g^0,\bar\sigma,\Phi]=
\sqrt{|g^0|}\Bigl\{
- \hat G_{AB}g^{0\mu\nu} \p_\mu\bar\sigma^A\p_\nu\bar\sigma^B \nn
-\sum_{s\in S} \eps_s \e^{2L_{As} \bar\sigma^A}
g^{0 \mu\nu} \p_\mu \Phi^s \p_\nu \Phi^s \Bigr \}.
\eeq

{\bf Proposition 3.} Let us consider the action \rf{7.1}, with the
potential \rf{7.5} satisfying orthogonality relations \rf{7.6}. Let metric
$g^0$ and $\hat\sigma=(\hat\sigma^A(x))$ satisfy equations of motion for
the action \rf{7.7} and the constraints imposed:  \beq{7.11}
L_{As}\hat\sigma^A=0,\quad s\in S.
\eeq

Let $g^0$, $\bar\sigma=(\bar\sigma^A(x))$ and $\Phi=(\Phi^s(x))$ satisfy
the  equations of motion for the action \rf{7.7a} and
\beq{7.12}
\bar\sigma^A=L^A{}_sf^s,
\eeq
where $f^s=f^s(x)$ are some functions, $L^A{}_s=\hat G^{AB}L_{Bs}$, $(\hat
G^{AB})=(\hat G_{AB})^{-1}$, $s\in S$. Then, the field configuration
\beq{7.13}
g^0,\quad \sigma=\hat\sigma+\bar\sigma,\quad \Phi
\eeq
satisfies the equations of motion \rf{7.2}--\rf{7.4}.

{\bf Proof.} The proposition can be readily verified using the relations
\bear{7.14}
\hat G_{AB}\p_\mu\bar\sigma^A\p_\nu\hat\sigma^B=0;
\\
\label{7.15}
V(\hat\sigma+\bar\sigma)=V(\hat\sigma),\quad
\frac\p{\p\sigma^A}V(\hat\sigma+\bar\sigma)=
\frac\p{\p\sigma^A}V(\hat\sigma); \\ \label{7.16}
L_{As}(\bar\sigma^A+\hat\sigma^A)=L_{As}\bar\sigma^A,
\ear
following from the conditions of Proposition 3.

Thus, we may find the exact solutions by two steps.
First, we should solve the equations of motion for the "truncated" model
\rf{7.7} and find "background" $(\hat\sigma,g^0)$ satisfying \rf{7.11}. On
the second stage we should solve the equations of motions
corresponding to \rf{7.7a} for the fields
$\bar\sigma$ and $\Phi$ on $(M_0,g^0)$-background with the
restriction of vanishing of total energy-momentum tensor for
$(\bar\sigma,\Phi)$-fields.

\subsection{Generalized intersecting $p$-brane solutions with
non-Ricci-flat spaces}

Here we apply the scheme considered in Subsect. 7.1
to the model \rf{2.1} with $\Lambda =0$.
Now the manifold is
\beq{7.17}
 M =M_0\times M_1\times\dots
\times M_n\times M_{n+1}\times\dots\times M_{n+k}
\eeq
instead of \rf{2.10} and the metric
\beq{7.18}
g=\e^{2\gamma(x)}g^0+\sum_{i=1}^{n+k}\e^{2\phi^i(x)}g^i
\eeq
instead of \rf{2.11}.

All $(M_i,g^i)$ are Einstein spaces, satisfying \rf{2.13},
$i=1,\dots,n+k$, with
\beq{7.19}
\xi_1=\dots= \xi_n=0,\quad \xi_{n+1} \ne 0,
\dots, \xi_{n+k} \ne 0.
\eeq

Then for electro-magnetic $p$-brane ansatz from Section 6
we get  according to Proposition 2
the $\sigma$-model \rf{7.1} with midisupermetric \rf{3.6}; $S$,
$\eps_s$ and $L_{As}$ are defined in \rf{6.30}, \rf{3.e}, \rf{5.17} and
\rf{3.7}, \rf{5.18} respectively and $i,j=1,\dots,n+k$.

The potential $V(\sigma)$, $\sigma=(\phi^i,\varphi^\alpha)$, in this case
has the form \rf{7.5} with
\bear{7.20}
u_{c i}=- 2 \delta^{n+ c}_i + \frac{2 d_i}{2-d_0},
\qquad  u_{c \alpha} = 0,    \\ \label{7.21}
A_c = - \frac12 \xi_{n+c} d_{n+c},
\ear
$c= 1,\dots,k$; $i=1,\dots,n+k$; $\alpha= 1,\dots,l$.

It may be verified that the vectors $u_c = (u_{cA})$  from
\rf{7.20} satisfy the orthogonality condition \rf{7.6}.
Indeed, the calculation gives for $s =(a,o,I)$ ($o = e,m$ and
$I \in \Omega_{a,o}$)
\beq{7.20e}
\hat G^{AB} u_{c A} L_{Bs}=
2 \frac{d(\{n +c \} \cap I)}{d_{n +c}} =0,
\eeq
since $\{n +c \} \cap I = \emptyset$  for
$c=1,\dots,k$ and  $I \in \{1, \ldots,n \}$.

Here $\Omega=\Omega(n)$
is unchanged, so all $p$-branes do not "live" in
non-Ricci-flat  "internal" spaces
$(M_{n+c},g^{n+c})$, $c=1,\dots,k$.

Then from Propositions 1, 3 and the results of Section 6 we obtain new
exact solutions with the metric
\beq{7.22}
g= U \left\{ \e^{2\hat\gamma(x)} g^0+
\sum_{i=1}^{n+k}U_i\e^{2\hat\phi^i(x)}g^i\right\}
\eeq
instead of \rf{6.31} and scalar field
\beq{7.23}
\varphi^\beta=\hat\varphi^\beta+\varphi_e^\beta+\varphi_m^\beta
\eeq
instead of \rf{6.33}. In \rf{7.22} $U$, $U_i$, $i=1,\dots,n$, are defined in
\rf{6.32} (here $D=\sum_{i=0}^{n+k}d_i$), $U_{n+1}=\dots=U_{n+k}=1$, and
\beq{7.24}
\hat\gamma= \gamma_0 (\hat \phi) =
\frac1{2-d_0} \sum_{i=1}^{n+k}d_i \hat\phi^i.
\eeq

The background fields $g^0$ and
$(\hat\sigma^A) = (\hat\phi^i(x), \hat\varphi^\alpha(x))$ satisfy the
equations of motion for the $\sigma$-model
\rf{7.7}  with $(G_{AB})$  defined in \rf{3.6} and
$(G_{ij})$ in (\ref{3.3}), $i, j = 1, \ldots, n+k$,
and
\beq{7.24b}
{V}(\hat \sigma) =
-\frac{1}{2}   \sum_{i =n+1}^{n+k} \xi_i d_i e^{-2 \hat\phi^i
+ 2 {\gamma_0}(\hat \phi)}.
\eeq

In other words the metric
\beq{7.25}
\hat g= \e^{2\hat\gamma(x)} g^0+\sum_{i=1}^{n+k}
\e^{2\hat\phi^i(x)}g^i
\eeq
and the set of scalar fields $\hat\varphi = (\hat\varphi^{\beta}(x))$
should satisfy the equations of motion for the action
\rf{2.1}  with $\Lambda=0$ and $F^a=0$, $a\in\tri$.
Background fields should also satisfy
the constraints (following from \rf{7.11})
\bear{7.26}
\lambda_a(\hat\varphi)-\sum_{i\in I}d_i\hat\phi^i=0,\quad
I\in\Omega_{a,e}, \quad a\in\tri_e, \\ \label{7.27}
-\lambda_b(\hat\varphi)-\sum_{j\in J}d_j\hat\phi^j=0,\quad
J\in\Omega_{b,m},  \quad b\in\tri_m.
\ear

Relations \rf{7.22}, \rf{7.23}, \rf{7.26}, \rf{7.27} are the only
modifications of the solutions from Section 6. (All other relations for
$F^a$, $\nu_s$, {\dots} are unchanged).

\section{Concluding remarks}

Using $\sigma$-model approach we have obtained
generalized composite electro-magnetic p-brane solutions.
The solutions (\ref{4.1}) with flat spaces
$(M_{\nu}, g^{\nu})$ one of which being pseudo-Euclidean one:
$M_{\nu} = {\bf R}^{d_{\nu}}$,
$g^{0} = \delta_{\mu \nu} dx^{\mu} \otimes dx^{\nu}$,
$g^{1} =
\eta_{m_{1} n_{1}}
dy_1^{m_{1}} \otimes dy_1^{n_{1}}$,
$g^i  = \delta_{m_{i} n_{i}} dy_i^{m_{i}} \otimes dy_i^{n_{i}}$ ($i > 1$)
and
\be
H_s(x) = 1 + \sum_{k = 1}^{N_s} \frac{ 2m_{sk}}{|x - x_{sk}|^{d_0 - 2}},
\qquad s \in S,
\ee
are usually interpreted in literature as intersecting p-branes
when all sets $I$ contain 1. In this case all p-branes
have common intersection containing $M_1$ manifold and
time submanifold belongs to  worldsheets of all p-branes.

Our solution may be considered as a generalization of intersecting
p-brane solutions to the case of
Ricci-flat (and also for some non-Ricci-flat) manifolds
$(M_{\nu}, g^{\nu})$ of arbitrary signatures. In this case
the submanifolds $M_I$ (\ref{2.20}) may be not intersecting
and may contain different time submanifolds, i.e. p-branes
may "live" in different times.
They may be considered as a starting point for a generalization of
multitemporal spherically-symmetric solutions \cite{IM5,IM6} to
the standard p-brane case.

\begin{center}
{\bf Acknowledgments}
\end{center}

This work was supported in part
by DFG grants 436 RUS 113/7, 436 RUS 113/236/O(R) and
by the Russian Ministry for
Science and Technology,  Russian Fund for Basic Research,
project N 95-02-05785-a. The authors
are grateful to Dr. M.Rainer
for his hospitality during their stay in Potsdam University
and to K.A.Bronnikov for valuable comments.

\end{document}